\documentstyle[11pt,subeqn,subeqnarray]{article}
\makeatletter
\def\abstract{\if@twocolumn
\section*{Abstract}
\else
\begin{center}
{\Large\bf Abstract\vspace{-.5em}\vspace{0pt}}
\end{center}
\quotation
\fi}
\def\endabstract{\if@twocolumn\else\endquotation\fi}
\makeatother
\renewcommand{\theequation}{\arabic{section}.\arabic{equation}}
\input {amssym.def}
\input {amssym.tex}
\newcommand{\req}[1]{(\ref{#1})}
\newcommand{\beq}[1]{\begin{equation}\label{#1}}
\newcommand{\eeq}{\end{equation}}
\title{\bf Metastability in Fluctuation Driven First-order
Transitions:  Nucleation of Lamellar Phases}
\author{{\em P. C. Hohenberg}\\
AT\&T Bell Laboratories\\
Murray Hill, NJ 07974 \bigskip \\
{\em J. B. Swift} \\
Department of Physics and Center for Nonlinear Dynamics \\
University of Texas, Austin, TX 78712}
\date{}

\begin{document}
\maketitle
\clearpage
\begin{abstract}

The nucleation of a lamellar phase from a supercooled
homogeneous phase in a fluctuation driven first-order
transition is studied, based on a phenomenological
free energy due to Brazovskii.
The absence of phase coexistence in the corresponding
mean-field approximation makes application of the standard
droplet theory of homogeneous nucleation problematic.
A self-consistent coarse-graining procedure
is introduced to overcome this difficulty, and the barrier height
for nucleation of a critical droplet is estimated in the
weak-coupling limit.
Contrary to earlier estimates the
critical droplet shape is shown to be anisotropic in general.
Some effects of distortions and defects in the lamellar structure
are considered and are shown to affect the critical
droplet only very near coexistence where the probability of
nucleation vanishes.
The coarse-graining procedure introduced here follows from
a novel application of the momentum-shell renormalization
group method to systems in the Brazovskii class.
Possible applications of the theory to the microphase
separation transition in diblock copolymers and to
Rayleigh-B\'{e}nard convection are briefly discussed.
\end{abstract}
\newpage
\setcounter{equation}{0}
\section{Introduction}
In 1975 Brazovskii [1] investigated isotropic or
nearly isotropic systems in which the fluctuation spectrum
had a minimum at a nonzero wavevector
$| {\bf q} | = q =q_0$, represented by a hypersphere
in $d$-dimensional reciprocal space.
He found that in the symmetric case where mean-field theory
would predict a continuous transition to a periodic ordered
state with spatial period $2\pi / q_0$, fluctuation
effects lead to a discontinuous or
{\it first-order} transition.
In the case of weak fluctuations (small noise), the
transition point is close to its mean-field value and
the self-consistent Hartree approximation employed by
Brazovskii could be justified as the leading term in a
systematic perturbation expansion.
The physical origin of the effect, which is essentially
independent of system dimensionality for
$d \geq 2$, lies in the large phase space
for one-dimensional fluctuations in the direction
transverse to the hypersphere.

Although initially proposed to describe weakly anisotropic
antiferromagnets and cholesteric liquid crystals, the
Brazovskii model was subsequently shown to apply to the
nematic to smectic-$C$ transition in liquid crystals [2],
to pion condensates in neutron stars [3], to the onset of
Rayleigh-B\'{e}nard convection [4,5] and most notably to the
microphase separation transition in symmetric diblock copolymers [6-9].
It is in this last system that experimental confirmation of the theory
was achieved [8] by quantitatively estimating  the mean-field
parameters and showing that the observed first-order transition
was inconsistent with
the predictions of mean-field theory, but
was well described by the Brazovskii theory.

Given this measure of success it is natural to ask about
the lifetime of the metastable isotropic phase as the system
is cooled below the thermodynamic transition point [10].
 From a theoretical point of view the problem is formulated by
classical homogeneous nucleation theory [11]
as well as by more sophisticated field-theoretic approaches [11, 12],
in terms of the free energy barrier for
creation of a droplet of ordered phase inside the disordered phase.
The critical droplet is the one which balances
the free-energy cost of the interface between coexisting
phases with the gain from the bulk ordering.
Mathematically, the barrier is calculated by first
solving a saddle-point or mean-field equation with the
boundary conditions that the system is disordered at
infinity and ordered at the origin, say.
Now for {\it fluctuation driven first-order transitions}
quite generally [13], there is no phase coexistence in mean-field theory
so this first step of the calculation cannot be taken.

Our work addresses this issue by developing a coarse-graining
procedure whereby the modes with wavevectors in the range
\begin{subequations}
\begin{equation}
\slabel{eq101a}
0 < q < q_0 - \Lambda , \;\; and \;\; q_0 + \Lambda < q < \infty
\end{equation}
are averaged over, and the remaining modes
with
\begin{equation}
\slabel{eq101b}
| q-q_0| < \Lambda
\end{equation}
\end{subequations}
are retained as fluctuating degrees of freedom.
When $\Lambda \rightarrow 0$ all modes are
averaged over and we recover
the bulk thermodynamic averages
calculated by Brazovskii [1].
For $\Lambda > 0$, we obtain an effective free
energy ${\cal F}_{\Lambda}$ for modes with wavevectors in the band
(1.1b), whose average wavevector is $q_0$ but
whose envelope can vary on a scale $L > \Lambda^{-1}$.
For $\Lambda$ not too large the bulk
phases described by ${\cal F}_{\Lambda}$ do show phase
coexistence in mean-field theory, so a
critical barrier height can be estimated from
classical nucleation theory for given
$\Lambda$.
We are then left with the problem
of determining the proper value of the
cutoff $\Lambda$, and for this we propose
a self-consistent procedure whereby $\Lambda$ is
equal to the local rate of variation of the
envelope in the critical droplet solution.

In cases when the ordered state is spatially anisotropic
it is important to optimize the {\it shape} of the critical
droplet, which will not be spherical due to the anisotropy
of the interface free energy.
The standard way to carry out this optimization is known
as a Wulff construction [14] and it arises in the present
problem as well.
The critical droplet we are led to for a lamellar
ordered state has the form of a long needle with
lamellae transverse to the needle axis.
In addition, just as for ordering in liquid crystals [15],
we find that under some circumstances line defects can be
introduced into the structure and lead to deformations of the lamellae
which lower the critical barrier height.
Since these effects depend on gaining surface free energy at the
cost of line energy,
defects are only favorable very near coexistence where
the critical droplet has large dimensions and the barrier height itself
is large.

In the Brazovskii theory there is a unique small parameter $\lambda$
representing the smallness of the dimensionless noise
strength, or equivalently of the dimensionless coupling constant.
The reduced temperature $\bar{\tau} = \bar{\tau}_c < 0$ at
which the first-order transition takes place is of order
$| \bar{\tau}_c | \sim \lambda^{2/3}$ below the mean-field
transition which takes place at $\bar{\tau} =0$.
For a $d$-dimensional system the critical free-energy
barrier for creation of an anisotropic
(Wulff) droplet is found to be
\begin{subeqnarray}
\slabel{eq102a}
\bar{B}_W &\sim&
\lambda^{(1-d)/6}
\left [ \frac{| \bar{\tau}_c |}{| \bar{\tau}|- |\bar{\tau}_c|} \right ]^{d-1}
, \; \; \; \;
| \bar{\tau} | \rightarrow | \bar{\tau}_c | , \\
\slabel{eq102b}
\bar{B}_W &\sim& \lambda^{(5-d)/2}
| \bar{\tau}|^{(d-7)/2} , \; \; \; \;
| \bar{\tau} | \gg | \bar{\tau}_c | \sim
\lambda^{2/3} .
\end{subeqnarray}
In contrast an isotropic droplet has a barrier
\begin{subeqnarray}
\slabel{eq103a}
\bar{B}_{iso} &\sim& \lambda^{(1-d)/3}
\left [ \frac{| \bar{\tau}_c |}{| \bar{\tau}|- |\bar{\tau}_c|} \right ]^{d-1}
, \; \; \; \;
| \bar{\tau} | \rightarrow | \bar{\tau}_c | , \\
\slabel{eq103b}
\bar{B}_{iso} &\sim& \lambda^{(3-d)}
| \bar{\tau} |^{d-4} , \; \; \; \;
| \bar{\tau} | \gg | \bar{\tau}| ,
\end{subeqnarray}
from which we see that the anisotropic barrier is always less than
the isotropic one.
An important conclusion of \req{eq102a} is that for
$| \bar{\tau} | \sim | \bar{\tau}_c | \sim \lambda^{2/3}$ the
dimensionless barrier
$[ \bar{B}_W \sim \lambda^{(1-d)/6} ]$
is still large for $\lambda \ll 1$ and $d \geq 2$, i.e. the
probability of nucleation is low.
On the other hand this is the point at which the droplet
size becomes of the same order as the interface width,
and for lower quenches ($| \bar{\tau} | > | \bar{\tau}_c |$) the
droplet will be ramified [16].
We thus do not expect spontaneous nucleation of well-defined
Wulff droplets.

Although our model and methods follow those of
Fredrickson and Binder [10] our conclusions are different.
These authors only estimated the barrier for isotropic droplets
in the domain \req{eq101a}, and did not consider either
ramified or anisotropic droplets, which according to
our estimates are the ones which are likely to be nucleated.

In Sec. II the coarse-grained free energy is obtained, first
by a phenomenological argument which follows the derivation
of the expanded Brazovskii free energy
in Ref. [10],
and then by using a novel  momentum-shell renormalization group.
In Sec. III interface and droplet free energies
are estimated, based first on the coarse-grained
free energy with fixed cutoff, and then on a
free energy with self-consistently determined cutoff.
The self-consistency is shown not to affect the scaling
of the nucleation barrier.
The anisotropy of the interface free energy on the other hand is
important, and it leads to an anisotropic critical droplet
whose shape is determined by a Wulff construction.
The contributions to the barrier height of defects and
distortions of the order inside the critical droplet
are estimated and it is shown that these effects are negligible
except very close to coexistence, when the nucleation barrier
and critical droplet dimensions are very large.
Section III concludes with a brief discussion of experiment
and of ways to pursue the theory in more quantitative
directions.
The derivations of the coarse-grained free energies are
described in the Appendixes.
\section{The Coarse-grained Brazovskii Model}
\subsection{Bulk behavior}
Our starting point is a phenomenological model
with relaxational (Model A) dynamics and a
Brazovskii free energy which we write as
\setcounter{equation}{0}
\begin{subeqnarray}
\slabel{eq201a}
\partial_{\bar{t}} \bar{\psi} & = & -
\frac{\delta \bar{\cal F}}{\delta \bar{\psi}} +
\bar{\eta} \;, \\
\slabel{eq201b}
\bar{{\cal F}} & = & \int d^d \bar{\bf {x}}
\left \{ \frac{1}{2}
\bar{\tau} \bar{\psi}^2 +
\frac{\lambda}{4!} \bar{\psi}^4 +
\frac{\tilde{\xi}_0^4}{2}
[ ( \bar{\bf {\bigtriangledown}}^2 +
\bar{q}_0^2 ) \bar{\psi} ]^2 \right \} , \\
\slabel{eq201c}
  &  & \langle \bar{\eta} ( \bar{\bf{x}} , \bar{t} )
\bar{\eta} ( \bar{\bf {x}}^{\prime} , \bar{t}^{\prime} ) \rangle =
2 \: \delta^{(d)}
( \bar{\bf {x}} - \bar{\bf {x}}^{\prime} )
\delta ( \bar{t} - \bar{t}^{\prime} ) \:.
\end{subeqnarray}
In the above equations the quantities
$\bar{q}_0 , \tilde{\xi}_0$ are considered
to be of order unity and there is a single
small parameter
\beq{eq202}
\lambda \; \ll \; 1 \:,
\eeq
and a control parameter $\bar{\tau}$.
The static (long-time) solution has the bulk
free energy
\beq{eq203}
\bar{\Phi} ( \bar{\tau} , \: \lambda ) =-
\ln \; \left [ \langle e^{- \bar{\cal F }} \rangle \right ] \; ,
\eeq
where the angular brackets in \req{eq203} denote an
average over the Gaussian noise $\bar{\eta}$,
which can be represented by a functional integral
[see below; the effective temperature, or
noise strength, has been scaled to unity in
Eq. \req{eq201c}].
The relation between our model and various
physical systems can be recovered by referring
to Eq. (2) of Brazovskii [1],  Eqs. \req{eq203} and
(4.1) of Fredrickson and Binder [10],  and
Eqs. (A26-29) of Hohenberg and Swift [5].
The essential point is that the small coupling
constant $\lambda$ reflects the smallness of the noise
strength in the original systems.

According to the derivation of Brazovskii [1], for
small $\lambda$ the self-consistent propagator in the
disordered phase is obtained from the Hartree
diagram in Fig. 1a, as
\beq{eq204}
\bar{g}^{-1} ( \bar{q}) =
\bar{r} +
\tilde{\xi}_0^4
( \bar{q}^2 - \bar{q}_0^2 )^2 ,
\end{equation}
with
\beq{eq205}
\bar{r} = \bar{\tau} + \frac{\lambda}{2}
\int \frac{d^d \bar{q}}{(2 \pi )^d}
\frac{1}{\bar{r} + \tilde{\xi}_0^4 ( \bar{q}^2 - \bar{q}_0^2 )^2} \; .
\eeq
As shown below, the solution
$\bar{r} ( \bar{\tau})$ of Eq. \req{eq205}
remains positive for all $\bar{\tau}$, so the
linear instability of the disordered state,
signalled by the vanishing of
$\bar{r}$ in mean-field theory ($\lambda =0)$, has been
completely eliminated in the Hartree approximation.
For $\bar{\tau} < 0$ there is a competing
ordered solution with [1]
\beq{eq206}
\langle \bar{\psi} ( \bar{{\bf x}} ) \rangle =
\bar{A} e^{i {\bf \bar{q}}_0 \cdot {\bf \bar{x}}}
+ cc \; ,
\eeq
a propagator leading to
\beq{eq207}
\bar{r}_A = \bar{\tau} +
\frac{\lambda}{2} \int \frac{d^d \bar{q}}{(2 \pi )^d}
\frac{1}{\bar{r}_A + \tilde{\xi}_0^4 ( \bar{q}^2- \bar{q}_0^2 )}
+ \lambda | \bar{A} |^2 ,
\eeq
and a field $\bar{h}$ conjugate to
$\bar{\psi}$, given by
\beq{eq708}
\bar{h} = \bar{A} (
\bar{r}_A - \frac{1}{2}
\lambda | \bar{A} |^2 ) \; .
\eeq
The bulk free energy difference per unit volume
$\Delta \bar{\Phi}$ between the disordered and the
ordered states can be obtained from the relation [1]
\beq{eq209}
\Delta \bar{\Phi} = \int_0^{\bar{A}}
\frac{\partial \bar{\Phi}}{\partial \bar{A}^{\prime}}
d \bar{A}^{\prime} =
\int_0^{\bar{A}}
2 \bar{h} \: d \bar{A}^{\prime} =
\int_{\bar{r}}^{\bar{r}_A}
2 \bar{h} \frac{d \bar{A}^{\prime}}{d \bar{r}^{\prime}}
d \bar{r}^{\prime} .
\eeq
For small $\bar{r}, \bar{r}_A$ Eqs. \req{eq205}
and \req{eq207}
may be rewritten in the form
\begin{eqnarray}
\label{eq2010}
\bar{r} =\bar{\tau} &+& \bar{\alpha} \lambda / \sqrt{\bar{r}} , \\
\label{eq2011}
\bar{r}_A = \bar{\tau} &+&
\bar{\alpha} \lambda / \sqrt{\bar{r}_A} + \lambda | \bar{A} |^2 ,
\end{eqnarray}
where
\beq{eq2012}
\bar{\alpha} =
( \bar{q}_0^{d-2} S_d \pi ) / 4 (2 \pi )^d
\tilde{\xi}_0^2 \equiv \pi \alpha / 2 \: ,
\eeq
and $S_d$ is the surface area of the $d$-dimensional unit sphere.
The essential point is that in all $d \gtrsim 2$, the integral
in Eqs. \req{eq205} and \req{eq207} contributes only near the surface
of the
$d$-dimensional sphere of radius
$\bar{q}_0$, and it is a {\it one-dimensional} integral in
the radial direction.
The transverse dimensions only contribute to the coefficient
$\bar{\alpha}$.
As shown by Brazovskii, the free energy difference
per unit volume $\Delta \bar{\Phi}$ of Eq. \req{eq209} then
takes the form
\beq{eq2013}
\Delta \bar{\Phi} = -
\frac{\bar{r}^2}{2 \lambda} -
\bar{\alpha} \bar{r}^{1/2} +
\frac{\bar{r}_A^2}{2 \lambda} +
\bar{\alpha} \bar{r}_A^{1/2} +
\frac{1}{4} \lambda | \bar{A} |^4 \: ,
\eeq
where  $\bar{r} ( \bar{\tau})$ and
$\bar{r}_A ( \bar{\tau})$ are given by \req{eq2010} and
\req{eq2011} [as mentioned in HS [5]  some numerical coefficients which were
incorrect in Eq. (14) of [1] have been changed].
The free energy difference changes sign at
\beq{eq2014}
\bar{\tau} = \bar{\tau}_c = -
2.03 ( \bar{\alpha} \lambda )^{2/3} \; ,
\eeq
which is the bulk (first-order) transition point in the
Hartree approximation.
Let us introduce the reduced variables
\begin{subeqnarray}
\slabel{eq2015a}
r & = & ( \alpha \lambda)^{-2/3} \bar{r} , \\
\slabel{eq2015b}
\tau & = & ( \alpha \lambda )^{-2/3} \bar{\tau} , \\
\slabel{eq2015c}
\psi & = & \lambda^{1/2} ( \alpha \lambda)^{-1/3} \bar{\psi} , \\
\slabel{eq2015d}
\langle \psi \rangle & = & A \: e^{i {\bf q}_0 \cdot {\bf x}} + cc \: , \\
\slabel{eq2015e}
A & = & \lambda^{1/2} ( \alpha \lambda )^{-1/3} \bar{A} , \\
\slabel{eq2015f}
{\bf x} & = & \bar{{\bf x}} / \bar{\ell} , \\
\slabel{eq2015g}
q_0 & = & \bar{q}_0 \bar{\ell} , \\
\slabel{eq2015h}
\bar{\ell} & = & (2 \tilde{\xi}_0^2 \bar{q}_0) ( \alpha \lambda)^{-1/3}
\equiv \bar{\xi}_0 ( \alpha \lambda )^{-1/3} ,
\end{subeqnarray}
in terms of which Eqs. \req{eq201b},
\req{eq205}, \req{eq207}, \req{eq2013} and \req{eq2014}
become
\begin{eqnarray}
\slabel{eq2016}
\bar{{\cal F}} & = & \beta {\cal F} , \\
\slabel{eq2017}
\beta & = & \bar{\ell}^d ( \alpha \lambda)^{4/3} \lambda^{-1} , \\
\slabel{eq2018}
{\cal F} & = & \int d^d x
\left \{ \frac{1}{2} \tau \psi^2 +
\frac{1}{4!} \psi^4 +
\frac{1}{2} (4q_0^2)^{-1}
[( \bigtriangledown^2+q_0^2) \psi ]^2 \right\} , \\
\slabel{eq2019}
r & = & \tau + \pi / (2r^{1/2} ) , \\
\slabel{eq2020}
r_A & = & \tau +
\pi / (2r_A^{1/2}) - \frac{1}{4} |A|^4 , \\
\slabel{eq2021}
\Delta \bar{\Phi} & = & \frac{(\alpha \lambda)^{4/3}}{\lambda} \Delta
\Phi =
\frac{(\alpha \lambda)^{4/3}}{\lambda}
\left [ \frac{1}{2} ( r_A^2 -r^2 ) +
\frac{\pi}{2} (r_A^{1/2} -r^{1/2} ) -
\frac{1}{4} | A |^4 \right ] \: ,\\
\slabel{eq2022}
\tau_c & = & -2.03 \:
( \pi / 2)^{2/3} = -2.74 .
\end{eqnarray}
In the scaling of Eq. (2.15) the coupling constant in
\req{eq2018} is of order unity and the only small parameter
appears in the coefficient of the gradient,
$q_0^{-2} \sim \lambda^{2/3}$.

It is instructive to expand the free energy difference $ \Delta \Phi$ in
the order parameter $| A |^2$, retaining only the first
three terms.
The Hartree result (2.21) can be considered as a function
of the independent variables $\tau$ and $| A|^2$ via Eqs. (2.19-20)
and expanded in the form
\beq{eq2023}
\Delta \Phi =  r |A|^2 +
\frac{1}{4} u |A|^4 +
\frac{1}{36} w |A|^6 +
O (|A|^8) .
\eeq
with $r=r ( \tau )$ given by the solution of Eq. \req{eq2019}, and
\begin{eqnarray}
\slabel{eq2024}
u & = & (1- \pi /4r^{3/2}) /(1+ \pi /4r^{3/2} ), \\
\slabel{eq2025}
w & = & (9 \pi /4r^{5/2}) / (1+ \pi /4r^{3/2} )^3 .
\end{eqnarray}
These functions are plotted in Fig. 2, from which it is
seen that $u$ becomes negative for $\tau < 0$, thus creating the
first-order transition.

The approximate free energy difference \req{eq2023} vanishes for
\beq{eq2026}
r = r_c = 9 u^2 / 16 w ,
\eeq
which occurs for
\beq{eq2027}
\tau =  \tau_c = - 2.51 ,
\eeq
which is close to the ``exact'' Hartree value
$\tau_c = - 2.74$ in Eq. \req{eq2022}.

Equation \req{eq2023} is precisely the same (when
allowance is made for slightly different scalings) as
Eq. \req{eq2022} of Fredrickson and Binder [10], which was arrived
at by using a ``Hartree potential''
$\Gamma_H [ \langle \psi \rangle ]$ in place of the ``bare''
free-energy functional \req{eq2018}.
While we believe that the proper physical interpretation
of this potential is in terms of a partially
coarse-grained potential ${\cal F}_{\Lambda} (\psi )$ (as explained in the
next section), for the purpose of calculating the bulk thermodynamic
properties the coarse-graining can be carried out to arbitrarily
long wavelengths ($\Lambda = 0$), and the potential
$\Delta \Phi$ of Eq. \req{eq2023} agrees with
$\Gamma_H [ \langle \psi \rangle ]$ of Ref. [10].

Let us inquire into the domain of validity of the
Hartree approximation \req{eq2021}.
As mentioned by Brazovskii [1] and by
Swift and Hohenberg [4], a simple estimate is obtained
by finding the parameter region where the correction term in
the self-energy [Fig. 1b] becomes of the same order as the
terms retained in Eq. \req{eq205} [Fig. 1a].
Let us write Eq. \req{eq205} generally as
\begin{eqnarray}
\slabel{eq2028}
\bar{r} & = & \bar{\tau} + \Sigma , \\
\slabel{eq2029}
\Sigma & = & \Sigma_H + \Sigma_2 + O ( \lambda^3) ,
\end{eqnarray}
with $\Sigma_H \sim \lambda / \bar{r}^{1/2}$ the term
retained in Eq. \req{eq205}, and
\beq{eq2030}
\Sigma_2 \sim \lambda^2 / \bar{r}^{3/2} ,
\eeq
the contribution from the diagram in Fig. 1b.
In order for the perturbation expansion in $\lambda$ to be valid we require
\beq{eq2031}
\Sigma_2 \sim \lambda^2 / | \bar{r} |^{3/2} \lesssim \bar{r} ,
\eeq
leading to $\bar{r} \gtrsim \lambda^{4/5}$.
In this domain the Hartree $\bar{r}$, given by
Eq. \req{eq2010}, scales as $\bar{r} \sim (\lambda / | \bar{\tau} |) ^2$
(i.e. $\bar{r} \ll | \bar{\tau}|$), and
Eq. \req{eq2031} implies
\beq{eq2032}
| \bar{\tau} | \lesssim | \bar{\tau}_{G1} | \sim \lambda^{3/5} .
\eeq
We use the subscript $G$ in \req{eq2032} in analogy to the
Ginzburg criterion for validity of mean-field theory
in critical phenomena [18].
Here, since we are dealing with a {\it first-order} transition
the theory is self-consistent so long as
\beq{eq2033}
| \bar{\tau}_c | \ll | \bar{\tau}_{G1} | ,
\eeq
which holds for
\beq{eq2034}
\lambda^{1/15} \ll 1 \: .
\eeq
We thus see that the strict self-consistency only
holds for impractically small coupling strengths.
We may also note that a less stringent criterion
was invoked by Brazovskii [1], namely
\beq{eq2035}
\Sigma_2 \lesssim \Sigma_1 = \Sigma_H ,
\eeq
which replaces Eq. \req{eq2032} with
\beq{eq3036}
| \bar{\tau} | \lesssim | \bar{\tau}_{G2} | \sim
\lambda^{1/2} ,
\eeq
and Eq. \req{eq2034} with
\beq{eq2037}
\lambda^{1/6} \ll 1 .
\eeq
\subsection{Nonuniform systems}
In order to study phase competition and
nucleation we must be able to describe situations
where the envelope $A$ of the order parameter
$\langle \psi \rangle$ of Eq. \req{eq2015d} can vary in space,
as for instance at the interface between ordered
and disordered domains.
For this purpose we introduce a
{\it coarse-graining procedure}, whereby modes with
wavevectors larger than some cutoff [19]  $ \Lambda$
are averaged over using the Hartree approximation
of Brazovskii, and modes with wavevector less than $\Lambda$ are
retained as fluctuating modes in averages such as Eq. \req{eq203}.
In particular, let us suppose that the starting free energy
\req{eq2018} involves modes with wavevectors in the range
$0 \leq q \leq 2 \Lambda_0 = 2q_0$, i.e.
\beq{eq2038}
0 \leq | q-q_0| \leq \Lambda_0 ,
\eeq
with $\Lambda_0 = q_0$.
The average in Eq. \req{eq203} is then a functional integral
\beq{eq2039}
\exp [- \beta \Phi (\tau) ] =
\int_{[0,\Lambda_0]}
{\cal D} [ \psi _q] \; \exp
\left \{ - \beta {\cal F} [ \psi_q ] \right \} ,
\eeq
where the symbol below the integral indicates the
inclusion of all modes $\psi(q)$ with wavevectors in
the range \req{eq2038} with upper cutoff $\Lambda_0$.
We now define a {\it coarse-grained} free energy ${\cal F}_{\Lambda}$
obtained by eliminating the modes in the slice
$\Lambda < | q-q_0| < \Lambda_0$.
This corresponds to setting
\beq{eq2040}
\exp [- \beta \Phi (\tau) ] =
\int_{[0,\Lambda]}
{\cal D} [ \psi_q] \:
\exp \{ - \beta {\cal F}_{\Lambda} [\psi_q] \} ,
\eeq
with
\beq{eq2041}
\exp [- \beta {\cal F}_{\Lambda} ] \equiv
\int_{[\Lambda , \Lambda_0]} {\cal D}
[ \psi_q] \; \exp
\{ - \beta {\cal F} [ \psi_q ] \} .
\eeq
With these definitions we show below that
\beq{eq2042}
\Phi = {\cal F}_{\Lambda =0} ,
\eeq
and
\beq{eq2043}
{\cal F} = {\cal F}_{\Lambda = \Lambda_0} .
\eeq
Moreover, in  Appendix B we show that if the integrals implied by
Eq. \req{eq2040} are
carried out using the Hartree approximation of
Brazovskii and if the resulting ${\cal F}_{\Lambda}$ is expanded in
$\psi^2$, then we obtain
\beq{eq2044}
{\cal F}_{\Lambda}
[ \psi ] =
\int_{[0, \Lambda ]} d^d {\bf x} \:
\left \{ \frac{1}{2} r ( \Lambda ) \psi^2 +
\frac{1}{4!} u ( \Lambda )
\psi^4 + \frac{1}{6!}
w ( \Lambda ) \psi^6
+ \left ( \frac{1}{4q_0^2} \right )
[( \bigtriangledown^2+q_0^2) \psi ]^2 \right \} ,
\eeq
where the symbol below the integral again indicates that
$\psi ( {\bf x})$ has variations involving wavevectors
in the range
$0 < |q-q_0| < \Lambda$.
The coefficients are given by [19]
\begin{eqnarray}
\slabel{eq2045}
r(\Lambda , \tau ) & = & \tau + \phi_1 ( \Lambda , \tau ) , \\
\slabel{eq2046}
u ( \Lambda , \tau ) & = &
[1- \phi_2 (\Lambda, \tau )] / [1+ \phi_2 ( \Lambda, \tau )], \\
\slabel{eq2047}
w ( \Lambda , \tau ) & = & 12 \phi_3 (\Lambda , \tau ) /
[1+ \phi_2 (\Lambda , \tau ) ]^3 ,
\end{eqnarray}
with [20]
\beq{eq2048}
\phi_n (\Lambda , \tau ) \equiv \int_{\Lambda}^{\infty}
\frac{dk}{[r( \Lambda , \tau ) + k^2]^n} .
\eeq
These quantities are evaluated by first solving the transcendental
equation \req{eq2045} for $r ( \Lambda , \tau )$ and then inserting
the result into $\phi_2$ and $\phi_3$ to obtain
$u( \Lambda , \tau )$ and $w ( \Lambda , \tau )$.

The derivation of Appendix B and the result in
Eqs. (2.45-48) are
closely related to those of Fredrickson and Binder [10], but the
physical content is rather different.
We are separating out the short-wavelength modes involved in
generating the first-order transition and the possibility of
phase coexistence $[u ( \Lambda ) < 0 ]$,
from the long-wavelength modes involved in building
interfaces and other large distortions of the order parameter.
The ``Hartree potential'' $\Gamma_H$ of
Fredrickson and Binder is essentially
${\cal F}_{\Lambda =0} = \Phi$,
which is no longer a functional of a fluctuating order parameter
$\psi (q)$ where $|q-q_0| < \Lambda$.
This potential depends only on the average order
parameter $ \langle \psi \rangle (q=q_0)$, which is only a function
of $\tau$ in a uniform system.
It thus seems to us that
${\cal F}_{\Lambda}$ rather than $\Gamma_H$ is the proper
starting point for the evaluation of droplet free
energies and metastability lifetimes, though as shown in the
next section our actual results do not differ
significantly from those obtained
using $\Gamma_H$.
To make a quantitative estimate of the droplet free
energy by means of ${\cal F}_{\Lambda} [ \psi ]$, Eq. \req{eq2044},
we must specify the value of $\Lambda$
and this is done in Sec. IIIB.

We should also mention that in principle
the coarse-graining procedure applies
to the full dynamics of Eq. \req{eq201a},
not just to the static averages such as Eq. \req{eq203}.
Since our treatment of metastability does not
go beyond the estimation of ``energy''
barriers [21], we have not pursued this
question here.
Clearly, however, a more complete theory would have to take into
account the effects of coarse-graining on the dynamics and on the
``entropic'' corrections to the barriers height coming from
fluctuations about the saddle point [11, 12].
\subsection{Properties of the coarse-grained free energy}
The model (2.44-48) defines a free energy
${\cal F}_{\Lambda}$ which was designed to
interpolate between the bare free energy ${\cal F}$ of
Eq. \req{eq2018} for [20]  $\Lambda = \Lambda_0 = \infty$,
and the bulk average free energy
$\Delta \Phi$ of Eq. \req{eq2023}
for $\Lambda =0$.
Indeed, it follows from Eq. \req{eq2048} that for
$\Lambda = \Lambda_0 = \infty$, $\phi_n ( \infty ) \equiv 0$,
so ${\cal F}_{\Lambda = \infty}$ just reproduces the bare free energy
${\cal F}$ of Eq. \req{eq2018}.
On the other hand from Eq. \req{eq2048} for $\Lambda =0$ we have
$\phi_1(0)= \pi /2 r^{1/2} (0)$,
$\phi_2(0)= \pi /4r^{3/2} (0)$,
$\phi_3= 3 \pi /16 r^{5/2} (0)$,
so $r(0)$, $u(0)$ and $w(0)$ agree with the quantities
$r, \; u ,\; w$, respectively, defined in
Eqs. \req{eq2019}, \req{eq2024} and \req{eq2025}.
Moreover, for $\Lambda =0$ there is only one mode left in
the free energy in Eq. \req{eq2044}, namely the average
order parameter, for which we make the Ansatz \req{eq2015d}.
The free energy per unit volume then becomes precisely the
expanded Hartree expression given in Eq. \req{eq2023}.

For intermediate $\Lambda$ values we may evaluate
$r( \Lambda , \tau )$, $u( \Lambda , \tau )$,
$w ( \Lambda , \tau )$ numerically from Eqs. (2.45-48).
The theory is of physical interest for values of $\Lambda$
which are low enough so that
$r ( \Lambda , \tau ) > 0$, implying that the
disordered phase is metastable.
In that case
it turns out that to good accuracy
we can represent the coefficients in the form
\begin{subeqnarray}
\slabel{eq2049a}
r ( \Lambda , \tau ) &\simeq& r(0,\tau) (1+\Lambda \tau) = r (1+
\Lambda \tau )  ,\\
\slabel{eq2049b}
u( \Lambda , \tau) &\simeq& u(0, \tau) = u ,\\
\slabel{eq2049c}
w( \Lambda ,\tau) &\simeq& w(0, \tau) = w ,
\end{subeqnarray}
results which are valid for
\beq{eq2050}
0 < \Lambda < - \tau^{-1} , \; \;  \; \; \tau < 0 .
\eeq
\subsection{A momentum-shell renormalization group}
It is natural to ask whether the
coarse-grained free energy obtained in the previous
section could not be derived in a more standard way,
using Wilson's momentum-shell renormalization group [22, 23], for
example.
It turns out that because the ordering in the Brazovskii
model \req{eq201b} involves condensation onto the surface of a
sphere in reciprocal space, the usual methods are difficult to
implement, and various
authors have found it necessary to introduce modifications of the
model in order to obtain renormalization group recursion relations [24,
25].
However, recent work on the renormalization group for
Fermi liquids at low temperatures, where the wavevectors of the important
modes also lie on a sphere, suggests that a
direct perturbation expansion might work.
Making use of some of the techniques developed
by Shankar [26] for the Fermi liquid we have been able to
derive recursion relations for the Brazovskii model \req{eq201b},
keeping essentially the same type of Hartree diagrams as in
earlier work.
The derivation is summarized in Appendix A, and the result is
again of the form \req{eq2044}, where now $r ( \Lambda)$, $u ( \Lambda)$
and $w ( \Lambda)$ are defined by
{\it differential recursion relations}
which are quoted in Eqs. (A.24).
Solutions of these equations yield
coefficients which are close to those of the simple
approximation (2.45-48)
for $\tau > 0$, and show a similar dependence
on $\Lambda$ and $\tau$ for $\tau < 0$.
There is, however, an important difference, in that the
solutions of the differential recursion relations (A.24)
are not defined for all $\Lambda$ and $\tau$ due
to singularities for $\tau < 0$, where $r(\Lambda , \tau) +
\Lambda^2=0$.
In particular the quantity
$r(0, \tau)$ vanishes at a finite
$\tau_1 < \tau_c$, unlike the solution
of \req{eq2045} which remains positive
for all $\tau$.
These singularities make it difficult to use the renormalization group
to estimate
droplet free energies for sufficiently negative $\tau$,
so we shall rather use the phenomenological
coefficients (2.45-48), which are
defined for all $\Lambda$ and $\tau$.
Nevertheless, the recursion relations are well behaved
for larger $\tau$ (including part of the metastable range), and they
are of intrinsic interest,
so we have presented their derivation in Appendix A.
\section{Droplet Theory of Nucleation}
\subsection{Phenomenological theory}
We shall be interested in estimating the free-energy barrier [21] for
nucleation
of a critical droplet of the ordered phase \req{eq2015d}
embedded in the disordered phase $\psi=0$ for
$\tau < \tau_c$.
Having obtained an effective free-energy ${\cal F}_{\Lambda}$
with $u < 0$, we first estimate the barrier
height for a critical droplet
with fixed $\Lambda$, deferring to the next subsection
the question of the proper choice of $\Lambda$.
As is done in standard nucleation theory [11-13], we
seek a saddle-point solution for ${\cal F}_{\Lambda}$, i.e. a
localized solution of the differential equation
\setcounter{equation}{0}
\beq{eq301}
\frac{1}{(4q_0^2 )} ( \bigtriangledown^2 +
q_0^2)^2 \psi +
r ( \Lambda) \psi +
\frac{1}{3!} u(\Lambda )
\psi^3 + \frac{1}{5!} w(\Lambda)
\psi^5 = 0 ,
\eeq
with $\psi=0$ for $| {\bf x}| \rightarrow \infty$
and $\psi \neq 0$ in the interior.
In this equation the coefficients $r, u, w$ are $O(1)$
and are given by Eqs. (2.46-49),
and $q_0 \sim \lambda^{-1/3} \gg 1$.
Given a solution $\psi_s$ of the above saddle-point equation,
the classical nucleation barrier [21] is
\beq{eq302}
\bar{B} = \bar{B}_{\Lambda}=
\beta {\cal F}_{\Lambda} [ \psi_s ] .
\eeq

Although we have now reduced the calculation to a straightforward
problem in differential equations, which could certainly be
attacked numerically, Eq. \req{eq301} is still quite difficult
to solve accurately, and we shall rather attempt to estimate
the value of $B_{\Lambda}$ analytically, in particular its dependence
on the parameters, which are $q_0$, $r$, $u$ and $w$ [in the
remainder of this subsection we consider these parameters to be basic, and
suppress their dependence on $\lambda$, $\Lambda$ and
$\tau$].
The difficulty in finding a general solution of Eq. \req{eq301}
arises from the existence of three widely different characteristic
lengths.
The first is the wavelength $q_0^{-1}$ of the ordered phase,
which in the present scaling (2.15) is vanishingly small.
Then there is the thickness of the interface between the
ordered and disordered phases, which is of order unity (see
below, however), and finally the droplet size, which diverges
at coexistence where the bulk free energies of the two
phases are equal.
\subsubsection{Interfaces}
We begin by considering only ordered states of the form
\beq{eq303}
\psi (x, {\bf y}) =
A(x, {\bf y})
e^{iq_0x} + c c ,
\eeq
describing small distortions of a pattern consisting of parallel
planes (or lines in two dimensions), with wavevector
in the $x$-direction, say.
Then for solutions whose envelope $A ( x, {\bf y})$ varies
slowly in space the latter can be shown to satisfy the
well-known amplitude equation [27]
\beq{eq304}
[ \partial_x + (i/2q_0) \partial_{{\bf y}}^2 ]^2
A (x, {\bf y} ) = 2r
A + u |A|^2A +
\frac{1}{6} w |A|^4 A ,
\eeq
where ${\bf y}$ here denotes the ($d-1$)-dimensional vector
transverse to the direction $x$ of alignment of the structure.
For bulk states, with $A = $constant, and for
$u < - (4r w /3)^{1/2}$, the equation
has two types of solutions, disordered
[$A=0$] and ordered [$A= \pm A_0$] with
\beq{eq305}
A_0^2 = 3 [( u^2 -4r w /3)^{1/2}
-u ] /w .
\eeq
Their free energies become equal when
$r=r_c$, with $r_c$ given by \req{eq2026}, though now
$r$, $u$, $w$ denote $r(\Lambda )$,
$u( \Lambda )$, $w ( \Lambda )$ rather than
$r(0)$, $u(0)$, $w(0)$, as in that equation.

Let us now consider {\it interfaces} between the ordered and disordered
states.
Clearly there are two types of simple interfaces, longitudinal
and transverse.
A {\it longitudinal} interface is a solution of
Eq. \req{eq304} in which $A$ only depends on $x$ with
$A =0$ for $x \rightarrow - \infty$, and
$A=A_0$ for $x \rightarrow + \infty$ say (see Fig. 3).
This is a quintic Ginzburg-Landau equation
as discussed, for example, by Fredrickson and
Binder [10]; the interface thickness is of order
\begin{subequations}
\begin{equation}
\slabel{eq306a}
\xi_{\parallel} \sim (2r)^{-1/2} ,
\end{equation}
and the interface free-energy (per unit area) is of order
\begin{equation}
\slabel{eq306b}
\sigma_{\parallel} \sim \xi_{\parallel} f_0 ,
\end{equation}
\end{subequations}
where $f_0$ is related to the bulk free energy difference
(per unit volume)
\setcounter{equation}{6}
\begin{eqnarray}
\label{eq307}
\Delta \Phi &=& -f_0 \delta , \\
\label{eq308}
\delta &=& (r_c -r ) / r_c , \\
\label{eq309}
f_0 &\simeq& 81 |u|^3/32 w^2 .
\end{eqnarray}
[If $f_0$ is defined by Eq. \req{eq309} then \req{eq307}
is strictly correct only for $\delta \rightarrow 0$, but it is
quite accurate up to $\delta =1$, and we shall use Eqs. (3.7-9)
for all $\delta$ in our estimates.]

A {\it transverse} interface, on the other hand
(Fig. 4), is a solution of
\beq{eq3010}
0 = \frac{1}{4q_0^2} \partial_y^4 A(y) +
2r A + u |A|^2 A +
\frac{1}{6} w |A|^4A ,
\eeq
which goes to $A_0$ for $y \rightarrow + \infty$ and to zero
for $y \rightarrow - \infty$, say.
The interface thickness is now of order
\beq{eq3011}
\xi_{\perp} \sim
[ 8q_0^2 r]^{-1/4} \sim \lambda^{1/6}
r^{-1/4} ,
\eeq
(which is small at fixed $r$ for $\lambda \ll 1$),
and the interface energy is
\beq{eq3012}
\sigma_{\perp} \sim \xi_{\perp} f_0 \ll \sigma_{\parallel} .
\eeq
It is this anisotropy in the interface energies which leads to
a preference for nonspherical droplets, as
detailed in the next subsections (this was missed by
Fredrickson and Binder [10]).
\subsubsection{Critical droplets: isotropic case}
Let us now estimate the free energy of a critical droplet solution of
Eq. \req{eq301}.
Near coexistence (i.e. for $\delta \ll 1$), the
dimensions of the droplet are very large,
and we can use the standard separation [12] of
${\cal F}$ into bulk and surface contributions to
estimate its free energy.
We first consider an isotropic droplet, i.e. one
made up of concentric lamellae (we refer to
three-dimensional structures in our discussion, but the results
are applicable in $d=2$ also).
Then the edge of the droplet is a purely
longitudinal interface [see Fig. 5a], and the
energy of a droplet of radius $R$ can be written in
the standard way as
\beq{eq3013}
\Delta {\cal F} (R) \simeq  R^{(d-1)} \sigma_{\parallel} -
R^d f_0 \delta ,
\eeq
The critical droplet is the one with maximum
free energy (as a function of $R$), namely
it has radius
\beq{eq3014}
R_{iso} \sim \xi_{\parallel} / \delta ,
\eeq
and corresponds to a free energy barrier
\beq{eq3015}
B_{iso} = \xi_{\parallel}^d f_0 / \delta^{d-1}
\sim [r^{-d/2} |u|^3 /w^2] / \delta^{d-1} .
\eeq
As usual the radius of the critical droplet and the
barrier height diverge at coexistence,
$r \rightarrow r_c$, $\delta \rightarrow 0$.
On the other hand, as the undercooling $\delta$
grows the radius decreases and when
$\delta = O(1)$ the droplet size $R$ is of the
same order as the interface thickness $\xi_{\parallel}$, and the
above
estimate base on a separation between bulk and surface free energies becomes
questionable.
As pointed out by Unger and Klein [16]
in an analogous situation, the
critical barrier
height may still be large at that point and it
is useful to extend the calculation to the ``ramified'' case
$\delta = O(1)$.

Let us first recast the above calculation in terms
of the free energy ${\cal F}_{\Lambda}$, Eq. \req{eq2044} and the
associated Euler-Lagrange equation \req{eq301}.
For an isotropic configuration made up of concentric
lamellae the wavevector of the ordered
state is radial, so the order parameter may be taken in the form
\beq{eq3016}
\psi = A ( \rho ) \; e^{iq_0\rho} + cc ,
\eeq
with real $A$, where $\rho$ is the radial coordinate (we neglect
all transverse variation for this simple estimate).
The equation for the droplet takes the form
\beq{eq3017}
-2 \partial_{\rho}^2 A - \frac{4}{\rho}
\partial_{\rho} A+2rA+u
A^3+ \frac{w}{6} A^5 =0 ,
\eeq
in leading order where terms neglected are of order
$( \rho q_0)^{-1} \partial_{\rho}^2A$,
$q_0^{-1} \partial_{\rho}^3A$, or higher.
The corresponding free energy is to leading order (in three dimensions)
\beq{eq3018}
{\cal F}_{iso} [A] = S_3 \int
d \rho \: \rho^2
\left\{ ( \partial_{\rho} A + \rho^{-1} A)^2+rA^2
+ \frac{u}{4} A^4+
\frac{w}{36}A^6 \right\} .
\eeq
Let us make a change of variables
\begin{subeqnarray}
\slabel{eq3019a}
A &=& (r/|u|)^{1/2} \tilde{A} , \\
\slabel{eq3019b}
\rho &=& r^{-1/2} \tilde{\rho} ,
\end{subeqnarray}
whereby \req{eq3017} and \req{eq3018} become,
for $u<0$
\begin{eqnarray}
\label{eq3020}
2 \partial_{\tilde{\rho}}^2 \tilde{A} &+&
(4 / \tilde{\rho}) \partial_{\tilde{\rho}}
\tilde{A} - 2 \tilde{A} +
\tilde{A}^3-
(3/32) (1- \delta )
\tilde{A}^5 =0 , \\
\label{eq3021}
{\cal F}_{iso} &=&
S_d r^{(4-d)/2} |u|^{-1}
\tilde{\cal F}_{iso} [ \tilde{A} ] ,
\end{eqnarray}
and for $d=3$
\beq{eq3022}
\tilde{\cal F}_{iso} [ \tilde{A}] =
\int d \tilde{\rho} \; \tilde{\rho}^2
\left \{ ( \partial_{\tilde{\rho}}
\tilde{A} + \tilde{\rho}^{-1} \tilde{A})^2 +
\tilde{A}^2 - \frac{1}{4}
\tilde{A}^4 +
(1/64) ( 1- \delta )
\tilde{A}^6 \right\} .
\eeq
where Eq. \req{eq308} was used.
[An analogous formula for $\tilde{\cal F}$ can be derived for arbitrary
$d$].

We now seek a (critical droplet) solution of Eq. \req{eq3020}
which vanishes for $\tilde{\rho} \rightarrow \infty$ and is
nonzero for $\tilde{\rho}=0$, with
$\partial_{\tilde{\rho}} \tilde{A} =0$ for
$\tilde{\rho} =0$.
The only parameter left
in Eqs. \req{eq3020} and \req{eq3022} is $\delta$,
and for $\delta \ll 1$ the solution reaches the bulk value
$\tilde{A}^2= \tilde{A}_0^2 = 8$ [see Eq. \req{eq305}] at
$\tilde{\rho}=0$.
The position $\tilde{R}$ of the interface results from a
balance between the surface and bulk terms in $\tilde{{\cal F}}$,
Eq. \req{eq3022}, and an argument
analogous to the one in Eq. \req{eq3013} leads to the results
${\bf \tilde{{\cal R}}} \sim \delta^{-1}$,
$\tilde{B}_{iso}= \tilde{{\cal F}}_{iso} [ \tilde{A}] \sim \delta^{-2}$
for $\delta \ll 1$.
 From Eq. \req{eq3021} we then see that
\beq{eq3023}
B_{iso}=r^{(4-d)/2}
|u|^{-1} \tilde{B}_{iso} \sim
r^{(4-d)/2} |u|^{-1}
\delta^{-(d-1)} ,
\eeq
which agrees with Eq. \req{eq3015} since for
$\delta \ll 1$, $r \approx r_c \sim |u|^2/w$.

Let us now consider the case $\delta =O(1)$ when
$\tilde{R} = O(1)$ and $\tilde{A}( \tilde{\rho} =0)$ no
longer reaches the bulk value
$\tilde{A}_0 ( \delta )$ corresponding to
\req{eq305}.
There is still expected to be
a unique solution of Eq. \req{eq3018}
satisfying the boundary conditions, and since all coefficients
in \req{eq3018} are of order unity, the barrier is given by
$\tilde{B}_{iso}= \tilde{{\cal F}}_{iso} [ \tilde{A} ]= O(1)$.
This corresponds to the ``ramified'' droplet of Unger and Klein [16],
for which the separation between bulk and surface is not applicable,
but a well-defined barrier can still be calculated from
Eq. \req{eq3020}.
Thus for arbitrary $\delta$ we have
\beq{eq3024}
B_{iso} \sim r^{(4-d)/2}
|u|^{-1} \tilde{B}_{iso} ( \delta ) ,
\eeq
with $\tilde{B}_{iso} ( \delta)$ an $O(1)$ function proportional
to $\delta^{1-d}$ for $\delta \ll 1$.
\subsubsection{Anisotropic droplets: the Wulff construction}
All of the above estimates were based on the assumption of an
isotropic solution \req{eq3014} of the saddle-point equation
\req{eq301}, with concentric lamellae (or rolls in $d=2$).
Since, however, the cost $\sigma_{\perp}$ of a transverse interface
is less than that of a longitudinal interface $\sigma_{\parallel}$
for small $\lambda$ [see Eq. \req{eq3012}], we
may guess that an anisotropic solution of \req{eq301} will
lead to a lower barrier.
The simplest such solution has parallel lamellae with
wavevector in the $x$-direction, say, as in
Eqs. \req{eq303} and \req{eq304}.

Let us first consider the classical nucleation regime
$\delta \ll 1$ and seek a solution of Eq. \req{eq301} which
vanishes for $x, | {\bf y} | \rightarrow \infty$
and reaches the bulk value \req{eq305} in the center
[see Fig. 5b].
Because of the anisotropy of the interface energy we must
optimize not only the size but also the {\it shape} of
the droplet, using the well-known Wulff construction [14].
In order to do this we consider a parallellipiped
of length $\gamma \sigma_{\parallel}$ in the $x$-direction
and $\gamma \sigma_{\perp}$ in the ($d-1$) transverse
(${\bf y}$)  directions, with $\gamma$ an unknown constant to be
determined (see Fig. 6).
The volume of the figure is
$\Omega= \gamma^d \sigma_{\perp}^{d-1} \sigma_{\parallel}$
so
\beq{eq3025}
\gamma = ( \Omega / \sigma_{\perp}^{d-1}
\sigma_{\parallel} )^{1/d} .
\eeq
Then the total free energy contribution from all the
surfaces is (see Fig. 6)
\begin{eqnarray}
{\cal F}_S &=& 2 \sigma_{\parallel}
(\gamma \sigma_{\perp} )^{d-1} +
2(d-1) \sigma_{\perp} (\gamma \sigma_{\parallel})
(\gamma \sigma_{\perp})^{d-1} , \nonumber \\
\label{eq3026}
   &=& 2d f_0 \Omega^{(d-1)/d}
(\xi_{\parallel} \xi_{\perp}^{d-1})^{1/d} ,
\end{eqnarray}
while the bulk contribution is
${\cal F}_B=-f_0 \Omega \delta$ [in deriving Eq. \req{eq3026},
Eqs. (3.6) and \req{eq3012} were used].
We now maximize the total energy
${\cal F} = {\cal F}_S + {\cal F}_B$ with
respect to the unknown volume $\Omega$, and find
the ``Wulff'' values
\beq{eq3027}
\Omega= \Omega_W \sim R_W^d ,
\eeq
with
\beq{eq3028}
R_W \sim \xi_W / \delta \sim
\sigma_W / f_0 \delta \sim
(\xi_{\parallel} \xi_{\perp}^{d-1} )^{1/d}  / \delta ,
\eeq
and a total free energy barrier
\beq{eq3029}
B_W \sim \xi_{\parallel} \xi_{\perp}^{d-1}
f_0 / \delta^{d-1} \sim \xi_W^d f_0 / \delta^{d-1} .
\eeq
The longitudinal and transverse dimensions are
\begin{subeqnarray}
\slabel{eq3030a}
(R_{\parallel} )_W & \sim &
\gamma_W \sigma_{\parallel} \sim \xi_{\parallel} / \delta , \\
\slabel{eq3030b}
( R_{\perp} )_W & \sim & \gamma_W \sigma_{\perp}
\sim \xi_{\perp} / \delta ,
\end{subeqnarray}
i.e. there is an anisotropy
\beq{eq3031}
(R_{\parallel})_W / (R_{\perp} )_W \sim q_0^{1/2}
\sim \lambda^{-1/6} ,
\eeq
which diverges for $\lambda \ll 1$.
Note also that the droplet becomes
ramified ($R_{\perp} \sim \xi_{\perp}$ and
$R_{\parallel} \sim \xi_{\parallel}$) for
$\delta \sim 1$, just as in the isotropic case \req{eq3014}.

In order to relate the above estimate to the
fundamental equation \req{eq301}, we note that for real $A$
Eq. \req{eq304} takes the form
\beq{eq3032}
\partial_x^2A+
(4q_0^2)^{-1}
\bigtriangledown_{{\bf y}}^4 A = 2rA+
uA^3+ \frac{1}{6} w A^5 .
\eeq
We seek a solution which vanishes for
$|x| , | {\bf y} | \rightarrow \infty$, and reaches
a nonzero value
over a finite region surrounding the origin.
Equation \req{eq3032} is derived from a free energy
\beq{eq3033}
{\cal F}_{ani} [A] =
\int dx d {\bf y}
\left\{ ( \partial_xA)^2+ \frac{1}{4q_0^2}
( \bigtriangledown_{{\bf y}}^2 A)^2
 + rA^2+ \frac{1}{4} uA^4+
\frac{1}{36} w A^6 \right\} .
\eeq
We now introduce the scaling of Eq. \req{eq3019a} and
\begin{subeqnarray}
\slabel{eq3034a}
x &=& r^{-1/2} \tilde{x} , \\
\slabel{eq3034b}
{\bf y} &=& r^{-1/4} ( 2q_0)^{-1/2} {\bf \tilde{y}} ,
\end{subeqnarray}
which changes Eqs. \req{eq3032} and \req{eq3033} into
\begin{eqnarray}\label{eq3035}
\partial_{\tilde{x}}^2 \tilde{A} &+& \partial_{\bf \tilde{y}}^4
\tilde{A} =
2 \tilde{A} - \tilde{A}^3 +
(3/32) (1- \delta) \tilde{A}^5 , \\
\label{eq3036}
{\cal F}_{ani} &=& r^{(7-d)/4} |u|^{-1}
(2q_0)^{(1-d)/2} \tilde{{\cal F}}_{ani} , \\
\label{eq3037}
\tilde{{\cal F}}_{ani} &=& \int d \tilde{x} d {\bf \tilde{y}}
\left\{ ( \partial_{\tilde{x}} \tilde{A})^2 +
(\partial_{{\bf y}}^2 \tilde{A})^2+
\tilde{A}^2 - \frac{1}{4}
\tilde{A}^4 + (1/64) (1- \delta )
\tilde{A}^2 \right\} .
\end{eqnarray}
In the classical nucleation region $\delta \ll 1$ the
solution of Eq. \req{eq3035} depends sensitively on $\delta$,
and the Wulff construction leading to Eq. \req{eq3029} tells
us that in the scaling of Eq. (3.34) we have
$\tilde{\cal F}_{ani} \sim \delta^{1-d}$,
$\tilde{R}_{\parallel} \sim \tilde{R}_{\perp} \sim \delta^{-1}$.
The solution becomes ramified
($\tilde{R}_{\parallel} \sim \tilde{R}_{\perp} \sim 1$) when
$\delta = O(1)$ and in that case once again
$\tilde{{\cal F}}_{ani} \sim O(1)$.
The free energy barrier \req{eq3029} may now be written as
\beq{eq3038}
B_W \sim \xi_{\parallel}
\xi_{\perp}^{d-1} f_0 / \delta^{d-1} \sim
q_0^{(1-d)/2} r^{-(1+d)/4}
|u|^3 / w^2 \delta^{d-1} \sim
q_0^{(1-d)/2} r^{(7-d)/4} / |u| \delta^{d-1} , \: \delta \ll 1 ,
\eeq
where we have again used the fact that for $\delta \ll 1$,
$r \sim r_c \sim |u|^2 / w$.
More generally we write, as in Eq. \req{eq3024},
\beq{eq3039}
B_W \sim q_0^{(1-d)/2} r^{(7-d)/4}
|u|^{-1} \tilde{B}_W ( \delta ),
\eeq
with
$\tilde{B}_W ( \delta ) =O(1)$,
$\tilde{B}_W \sim \delta^{1-d}$ for $\delta \ll 1$.
\subsection{Self-consistent theory}
Having estimated the barrier height for the
simplest isotropic and anisotropic critical droplets using the phenomenological
model \req{eq2044} with given $q_0$, $r$, $u$ and $w$, we
now address the issue of the proper choice of the coarse-graining scale
$\Lambda$.
The reason for introducing a coarse-grained free energy
in the first place was that we needed to allow for
variations of the order parameter on length scales of the
order of the interface width $\xi$ and larger.
Thus, in order to estimate the longitudinal surface
free-energy $\sigma_{\parallel}$ we need to impose
the constraint
\beq{eq3040}
\Lambda^{-1} \lesssim \xi_{\parallel} .
\eeq
It is apparent from Fig. 7 that if ${\bf q}$ is
confined to a shell of thickness $2 \Lambda$ around the point
$q_0 {\bf \hat{x}}$, then for $q_0 \gg 1$ the
transverse momentum $| {\bf q}_y | = q_y$ is limited by
$\Lambda_{\perp}= (2q_0 \Lambda)^{1/2}$.
Thus, since $\xi_{\perp}=(2q_0 \xi_{\parallel})^{1/2}$ the
constraint
\req{eq3040} automatically ensures that
\beq{eq3041}
\Lambda_{\perp}^{-1} \lesssim \xi_{\perp} ,
\eeq
which is also a necessary condition for evaluating
$\sigma_{\perp}$ consistently from
${\cal F}_{\Lambda}$.
In order to estimate bulk contributions to the droplet
free energy, on the other hand, we need to let
$\Lambda \rightarrow 0$ as in Eq. \req{eq2023}.
These different constraints can be implemented concurrently by using
a type of local-density or Thomas-Fermi approximation,
whereby $\Lambda$ is adjusted {\it self-consistently} to be equal
to the local longitudinal rate of variation
of the envelope $A(x, {\bf y} )$:
\beq{eq3042}
\Lambda = \Lambda_A (x) \equiv A^{-1} \partial_x A .
\eeq
We thus seek an extremum of the free energy functional
\beq{eq3043}
{\cal F}_{ani}^{sc} [A; \Lambda ] = \int dx d {\bf y}
\left\{ ( \partial_xA)^2+
\left ( \frac{1}{4q_0^2} \right )
( \bigtriangledown_{{\bf y}}^2 A)^2 +
+ r( \Lambda) A^2+ \frac{1}{4} u( \Lambda) A^4 +
\frac{1}{36} w (\Lambda) A^6 \right\} ,
\eeq
with $\Lambda$ determined self-consistently at each point
via Eq. \req{eq3042}.
It can be shown that the Euler-Lagrange equation is
no longer precisely given by Eq. \req{eq304}, but rather by
\beq{eq3044}
\partial_x^2A-(4q_0^2)^{-1}
\bigtriangledown_{{\bf y}}^4 A =
2 \hat{r}(\Lambda) A+ \hat{u} (\Lambda)A^3
+ \frac{1}{6} \hat{w}(\Lambda) A^5 ,
\eeq
with
\begin{subeqnarray}
\slabel{eq3045a}
\hat{r}(\Lambda) &=& r(\Lambda) + \Lambda dr/d\Lambda , \\
\slabel{eq3045b}
\hat{u} (\Lambda) &=& u(\Lambda)+ \Lambda du/ d\Lambda ,\\
\slabel{eq3045c}
\hat{w}( \Lambda ) &=& w(\Lambda) +
\Lambda dw/d\Lambda .
\end{subeqnarray}
Referring to Eqs. (2.49) we see that $du/d \Lambda$ and
$dw/d \Lambda$ are small, while $dr/d \Lambda \sim \tau$ so
we set
\begin{subeqnarray}
\slabel{eq3046a}
\hat{r} (\Lambda) &=& r(0) [1+2 \Lambda \tau ], \\
\slabel{eq3046b}
\hat{u}(\Lambda) &=& u(\Lambda )=u , \\
\slabel{eq3046c}
\hat{w}(\Lambda) &=& w(\Lambda) = w .
\end{subeqnarray}

The self-consistent droplet is given by the solution of
Eq. \req{eq3044},
subject to the constraint \req{eq3042} or
equivalently by the saddle point of \req{eq3043}, subject to the
same constraint.
For the isotropic droplet Eq. \req{eq3042} is replaced by
\beq{eq3047}
\Lambda_A ( \rho ) = A^{-1} \partial_{\rho} A,
\eeq
and the gradient terms in \req{eq3043} and \req{eq3044}
are replaced by $[\partial_{\rho}A+ (2 / \rho)A]^2$
and $[\partial_{\rho}^2 A+ (2/ \rho) \partial_{\rho}A ]$,
respectively, as in Eqs. \req{eq3017} and \req{eq3018}.

We have not solved these equations numerically,
but we may once again use dimensional arguments to estimate
the effect of the self-consistency on the barrier height.
In the isotropic case and for $\delta \ll 1$ we argue
that the main change compared to Eq. \req{eq3015} is
to select $\Lambda = \xi_{\parallel}^{-1}$ in evaluating the surface
term, whereas we may take $\Lambda= R^{-1} \sim \delta \sim 0$
in the bulk term.
Thus the barrier is changed to
\beq{eq3048}
B_{iso}^{sc} \sim \xi_{\parallel}
( \hat{r} (\xi_{\parallel}^{-1} )) f_0 ( \hat{r} (0)) / \delta^{d-1}.
\eeq
Now from Eq. \req{eq3046a} and the relation
$\xi_{\parallel} ( \hat{r}) \sim \hat{r}^{-1/2}$ we may show that
\beq{eq3049}
\xi_{\parallel} (\hat{r} ( \xi_{\parallel}^{-1})) \sim
3.2 \xi_{\parallel} ( \hat{r} (0)) = 3.2 \xi_{\parallel} ,
\eeq
i.e. the self-consistency increases the barrier by a
factor of order unity.
More generally, for arbitrary $\delta$, we introduce the scaling
(3.19), but where now $r$ and $u$ stand for
$r(\Lambda=0)$, $u(\Lambda=0)$.
Then Eqs. \req{eq3020} and \req{eq3022} become, respectively
\beq{eq3050}
2 \partial_{\tilde{\rho}}^2 \tilde{A} +
( 2 / \tilde{\rho} ) \partial_{\tilde{rho}} \tilde{A} -
2 \tilde{r} ( \tilde{\Lambda} ) \tilde{A} +
\tilde{A}^3 - (3/32) (1-\delta)
\tilde{A}^5=0 ,
\eeq
and
\beq{eq3051}
\tilde{\cal F}_{iso}^{sc} [ \tilde{A} ]
= \int d \tilde{\rho} \: \tilde{\rho}^2
\left\{ ( \partial_{\tilde{\rho}} \tilde{A} +
\tilde{\rho}^{-1} \tilde{A} )^2 +
\!\tilde{\tilde{r}} ( \tilde{\Lambda} )
\tilde{A}^2 - \frac{1}{4}
\tilde{A}^4+
(1/64) (1- \delta )
\tilde{A}^2 \right\} ,
\eeq
with
\begin{eqnarray}\label{eq3052}
\delta &=& [r(\Lambda=0)-r_c] /r_c =(r-r_c)/r_c , \\
\label{eq3053}
\tilde{r} ( \tilde{\Lambda} ) &=& 1+ \tilde{\Lambda} r^{1/2} \tau , \\
\label{eq3054}
\!\tilde{\tilde{r}} ( \tilde{\Lambda} ) &=& 1+ 2 \tilde{\Lambda}
r^{1/2} \tau .
\end{eqnarray}
The self-consistency relation \req{eq3047} becomes
\beq{eq3055}
\tilde{\Lambda} = \tilde{\Lambda}_{\tilde{A}}
( \tilde{\rho}) = \tilde{A}^{-1}
\partial_{\tilde{\rho}} \tilde{A} .
\eeq
Once again all coefficients in \req{eq3050} are of order
unity for $\delta=O(1)$, so Eq. \req{eq3023} becomes
\beq{eq3056}
B_{iso}^{sc} \sim r^{(4-d)/2}
|u|^{-1} \tilde{B}_{iso}^{sc} (\delta) ,
\eeq
where $r$, $u$ and $\delta$ refer to $\Lambda=0$,
and $\tilde{B}_{iso}^{sc}=O(1)$ is obtained by
solving \req{eq3050} subject to the constraint
\req{eq3055}.

In the anisotropic case the arguments go through in the
same way, and they lead to the replacement of \req{eq3029} for
$\delta \ll 1$, by
\beq{eq3057}
B_W^{sc} \sim \xi_{\parallel}
[r( \Lambda=\xi_{\parallel}^{-1} ) ]
\xi_{\perp}^{d-1} [r(\Lambda=\xi_{\parallel}^{-1} ) ]
f_0 [r( \Lambda= R_{\parallel}^{-1} )] \delta^{1-d} .
\eeq
Now for $\delta \ll 1$,
$R_{\parallel}^{-1} \sim \delta \rightarrow 0$,
and once again Eqs. (2.49) and \req{eq306a} lead to
\begin{subequations}
\begin{equation}
\slabel{eq5058a}
r(\Lambda=\xi_{\parallel}^{-1} ) = 0.1
r(\Lambda=0) .
\end{equation}
and
\begin{equation}
\slabel{eq3058b}
\xi_{\parallel} (\Lambda = \xi_{\parallel}^{-1} ) =
3.2 \; \xi_{\parallel} ( \Lambda =0) .
\end{equation}
\end{subequations}
Thus the self-consistent anisotropic barrier is increased
with respect to \req{eq3029} by a factor of order
unity, and Eq. \req{eq3038} still holds,
with $r$, $u$ and $\delta$ now referred to $\Lambda=0$.

For $\delta$ of order unity the scaling of Eqs. \req{eq3019a} and
(3.34) goes through with $r=r(\Lambda=0)$, and
Eqs. \req{eq3035} and \req{eq3037} are replaced by
\begin{eqnarray}
\label{eq3059}
\partial_{\tilde{x}}^2 \tilde{A} &+&
\partial_{\tilde {y}}^4 \tilde{A} \; \; \; = \; \; \; 2 \tilde{r} (
\tilde{\Lambda} )
\tilde{A} - \tilde{A}^3 + (3/32) (1-\delta) \tilde{A}^5 , \\
\label{eq3060}
\tilde{{\cal F}}_{ani}^{sc} [ \tilde{A}] &=& \int d \tilde{x}
d {\bf y}
\left\{ ( \partial_{\tilde{x}} \tilde{A} )^2 +
( \partial_{\bf y}^2 \tilde{A})^2 +
\tilde{\tilde{r}} ( \tilde{\Lambda} ) \tilde{A}^2 -
\frac{1}{4} \tilde{A}^4 +
(1/64) (1- \delta) \tilde{A}^6 \right\} ,
\end{eqnarray}
with the self-consistency relation
\beq{eq3061}
\tilde{\Lambda}= \tilde{\Lambda}_{\tilde{A}} ( \tilde{x}) =
\tilde{A}^{-1} \partial_{\tilde{x}} \tilde{A} ,
\eeq
and $\tilde{r} ( \tilde{\Lambda})$,
$\tilde{\tilde{r}} ( \tilde{\Lambda})$ given by Eqs. \req{eq3053}
and \req{eq3054}, respectively.
Then the self-consistent estimate of the anisotropic barrier
has the same form as Eq. \req{eq3039},
\beq{eq3062}
B_W^{sc} \sim q_0^{(1-d)/2} r^{(7-d)/4}
|u|^{-1} \tilde{B}_W^{sc} ( \delta ) ,
\eeq
but now $r$, $\delta$ and $u$ refer to
$\Lambda=0$, and $\tilde{B}_W^{sc} (\delta) =O (1)$
is obtained by solving the self-consistent equations
(3.59-61).

The final results may then be expressed in terms of
the original variables of Eq. \req{eq201b}, namely
$\bar{\tau}$ and $\lambda$, by using the formulas of Sec. IIA.
We defer a detailed examination of the results until
Sec IIID below, but it should already
be clear that the anisotropic barrier \req{eq3062} will always
be smaller than the isotropic one \req{eq3056}, due
to the factor $q_0^{(1-d)/2} \sim \lambda^{(d-1)/6} \ll 1$ and
the higher power of $r \lesssim 1$ in Eq. \req{eq3062}.
\subsection{Defects and distortions}
The calculation of the anisotropic barrier was based on the
simplest ansatz designed to take advantage of the
favorable transverse interface
energy $\sigma_{\perp}$, namely parallel
lamellae with wavevector
${\bf q} \simeq q_0 {\bf \hat{x}}$, as in Eq. \req{eq303}.
It should be clear, however, that under certain
circumstances a lower-energy configuration can
be achieved by distorting the lamellae in regions where they form
(unfavorable) longitudinal interfaces, in order to
gain additional surface free energy.
In general such distortions tend to produce defects,
whose free energy cost is linear in the size of the structure,
so distortions are typically favored for large droplets.

Our discussion of defects and distortions draws heavily on the
work of Cross [28] and of Fournier and Durand [15].
In particular, the latter authors showed that for smectic liquid
crystals it is advantageous to introduce defects in the form
of focal conics, and to pack these into
overall spherical shapes in the asymptotic limit of an
infinite droplet (Fig. 8).
As shown below, the same arguments hold for the present
system but it turns out that focal conic defects
are only favorable in a small region near coexistence
($\delta \ll 1$, $\tau \rightarrow \tau_c$, $R \rightarrow \infty$).
Similarly, we have considered overall distortions of the anisotropic
droplet into an annulus, in order to eliminate the
costly longitudinal interfaces at the tips (see Fig. 9).
This introduces dislocations in the {\it bulk} of the
structure, however, and according to our estimates it is
not favorable for any value of $\tau$.

Unless otherwise noted we will consider three-dimensional systems
in this section, though similar arguments can be given for
$d=2$, or any other $d>3$.
Following Cross [28] let us write the gradient term
in the free energy \req{eq2018} in the form
\beq{eq3063}
{\cal F}_G \equiv \frac{1}{8 q_0^2} \int
d^3 {\bf x}
\left [ ( \bigtriangledown^2+q_0^2) \psi \right ]^2 =
\frac{1}{2} \int d^3 {\bf x} \: \kappa
[( {\bf \bigtriangledown} \cdot {\bf \hat{n}})^2 +
4( \delta q)^2 ],
\eeq
where
\begin{eqnarray}
\label{eq3064}
\psi &=& Ae^{i\zeta} + cc, \\
\label{eq3065}
\mbox{\boldmath $ \bigtriangledown$} \zeta &=& (q_0+ \delta q) {\bf
\hat{n}} ,
\end{eqnarray}
and the bending constant is given by
\beq{eq3066}
\kappa = \frac{1}{2} A^2
\eeq
in these units.

Let us now consider the effect of inserting a focal
conic into the Wulff droplet, as sketched
in Fig. (8a,b),
and let us denote by $L$ the average radius of curvature
of the bend in the structure.
This length also corresponds to the length
of the core of the disclination defect line, which has
an average cross section $\xi_{\parallel}^2$.
Then as argued by Fournier and Durand [15], the
introduction of the focal conic yields a bulk energy cost
of bending of order $\kappa L^{-2}$ per unit volume (or
$\kappa L$ overall), and a term
$\kappa L \ln (L/ \xi_{\parallel} )$ to
account for the core of the defect.
The surface energy, on the other hand, is decreased by an
amount $\sim \Delta \sigma L^2$ where
\beq{eq3067}
\Delta \sigma = \sigma_{FC} - \sigma_W
\eeq
is the difference between the structures in Figs. 8b
and 5b, respectively.
Now according to the calculation in Sec. IIIA3 we
have $\sigma_W \sim f_0 ( \xi_{\parallel} \xi_{\perp}^{2})^{1/3}$,
whereas for the structure with a focal conic most of the
interface is transverse, so
$\sigma_{FC} \sim f_0 \xi_{\perp}^3 \ll \sigma_W$.
The free energy difference between the two
structures scales as
\beq{eq3068}
{\cal F}_W - {\cal F}_{FC} \sim
\kappa L [c_1+c_2 \ln (L/ \xi_{\parallel} )] +
\Delta \sigma L^2 ,
\eeq
where $c_1$ and $c_2$ are constants of order unity.
The above quantity vanishes for a size $L$ of order
\beq{eq3069}
L_0 = \kappa / | \Delta \sigma | \sim \kappa / \sigma_W ,
\eeq
which means that it is favorable to introduce
focal conics into the Wulff
droplets for $R_W = \sigma_W / f_0 \delta > L_0$, i.e. for
\beq{eq3070}
\delta < \delta_{FC} \sim
\frac{\sigma_W}{f_0L_0} \sim
\frac{\sigma_W^2}{f_0 \kappa} \sim
q_0^{-2/3} ,
\eeq
where we have used the relations
$\sigma_W \sim \xi_W \sim ( \xi_{\parallel} ) \xi_{\perp}^{1/3}$,
Eqs. \req{eq3066} and \req{eq3011}
and the fact that in the present scaling and near
coexistence ($\delta \ll 1$) all the coefficients
$r \sim r_c$, $u$, $w$,
$f_0$, $\kappa \sim A^2 \sim r/u$ are of order
unity and only factors of $\delta$ and
$q_0 \sim \lambda^{-1/3} \gg 1$ need to be
considered.
In the range \req{eq3070} the barrier is then determined
by $\sigma_{\perp}$, i.e. it scales as
\beq{eq3071}
B_{FC} \sim \xi_{\perp}^3 / \delta^2 \sim
q_0^{-3/2} / \delta^2, \; \; \; \;
\delta < q_0^{-2/3} ,
\eeq
rather than \req{eq3038}.

The above calculation assumed that the Wulff droplet was
unchanged in shape and size, and only the texture
of the lamellae was modified.
Following Ref. [15], we may consider
a composite droplet consisting of a spherical array of
conical domains, each one of which contains
a focal conic defect (see Fig. 8d).
Such a shape was found to predominate for
large droplets in the case of smectic liquid crystals [15].
The surface free energy is now still transverse and it is
achieved over a spherical surface
($\sigma_{\perp} R^2$), since the cost
is entirely in line energy
($\sim R \ln R$) and is negligible for large $R$.
Thus the free energy is given by
\beq{eq3072}
{\cal F}_{s ph} \sim \sigma_{\perp} R^2 -
\delta f_0R^3 ,
\eeq
leading to
\beq{eq3073}
R_{s ph} \sim \xi_{\perp} / \delta .
\eeq
This is more favorable than the single conic of
Fig. (8a,b) for $R_{s ph} > L_0$, i.e.
\beq{eq3074}
\delta < \delta_{s ph} =
\sigma_{\perp} / f_0L_0 \sim
\sigma_{\perp} \sigma_W \sim q_0^{-5/6} .
\eeq
The barrier for this droplet has the same scaling as the one with
a single focal conic, but presumably with a smaller
coefficient in the range \req{eq3074},
\beq{eq3075}
B_{s ph} \sim q_0^{-3/2} \delta^{-2} , \; \; \;
B_{s ph} < B_{FC} , \; \; \;
\delta < q_0^{-5/6} .
\eeq

Finally, let us estimate the free energy barrier for creation of an
annular droplet (Fig. 9), which lowers the surface
energy at the cost of splay and/or defect energy in
the bulk.
We first consider an undefected structure in which the
bending of the lamellae leads to a change in the local
wavevector, and a bulk energy cost given by the last
term in Eq. \req{eq3063}.
If we denote the radius by $R_1$ and the width by
$R_2 \ll R_1$, then the change in wavevector from
the inner to the outer rim is
$\delta q \sim q_0R_2 / R_1$, and the energy cost is
$( \delta q)^2 R_2^2 R_1 \sim q_0^2R_2^4 / R_1$.
The surface energy is entirely transverse, so the total energy
of the annulus is
\beq{eq3076}
{\cal F}_{annu} \sim q_0^2R_2^4 / R_1+
q_0^{-1/2} R_1R_2 -R_1R_2^2 \delta ,
\eeq
where near coexistence ($r \sim r_c$, $\delta \ll 1$) we
may again set $f_0 \sim 1$,
$\sigma_{\perp} \sim \xi_{\perp} \sim q_0^{-1/2}$.
In order to find the critical size and shape we
need to maximize \req{eq3076} at fixed ratio
\beq{eq3077}
\eta = R_2 /R_1 ,
\eeq
and then minimize the result with respect to the
ratio $\eta$.
We find $\eta_c \sim \delta^{1/2} q_0^{-1}$,
$R_{1c} \sim q_0^{1/2} \delta^{-3/2}$ and
a barrier
\beq{eq3078}
B_{annu} \sim q_0^{-1/2} \delta^{-5/2} ,
\eeq
which is always larger than the Wulff value
$B_W \sim \sigma_W / \delta^2 \sim \xi_{\perp}^2 / \delta^2 \sim 1/q_0
\delta^2$,
for $\delta \lesssim 1$.

Thus the cost of changing the local wavevector is too high so
we shall keep the constraint $q \approx q_0$ on
average, and attempt to achieve bend by introducing dislocations
in the bulk of the structure (Fig. 9b).
Since dislocations comes form lines in three dimensions
or points in two dimensions, the 2d estimate of Cross [28]
for the number $N_D$ of dislocations applies in
3d also, namely
\beq{eq3079}
N_D \sim q_0 R_2 .
\eeq
Each dislocation costs an energy
$f_0 \delta \xi_{\parallel} \xi_{\perp}$ per unit
length of core and has length $R_2$, so the bulk energy cost of
the dislocations is
$\sim f_0 \delta \xi_{\parallel} \xi_{\perp} R_2^2 q_0$.
Thus the structure in Fig. 9b has energy
\beq{eq3080}
{\cal F}_D \sim \delta q_0^{1/2} R_2^2 +
q_0^{-1/2} R_1R_2 -
R_1R_2^2 \delta .
\eeq
The same procedure as above then yields
$\eta_c \sim (q_0 \delta)^{-1}$,
$R_{1c} \sim q_0^{1/2}$, and a barrier
\beq{eq3081}
B_D \sim (q_0^{1/2} \delta)^{-1} .
\eeq
The above estimates are only valid if
$R_2<R_1$, i.e. $\eta_c<1$ which means
\beq{eq3082}
\delta > q_0^{-1} .
\eeq
In the next section we put together all the estimates to find the
most favorable structure for each region of $\delta$
(or $\tau$).
\subsection{Results}
In Eqs. \req{eq3056}, \req{eq3062}, \req{eq3075},
\req{eq3078} and \req{eq3081} we have
presented estimates for the barrier heights of
the droplets shown in Figs. (5a,b), (8b,d) and (9b),
expressed in the scaled units of Eq. (2.15),
as a function of $q_0 \sim \lambda^{-1/3}$ and the
bulk parameters $r$, $u$, $w$, $\delta=(r_c-r) /r_c$
(evaluated at $\Lambda=0$), which depend on $\tau$.
Near the bulk transition ($\delta \ll 1$) only the
parameters $\delta \ll 1$ and
$q_0 \sim \lambda^{-1/3} \gg 1$ survive since
$r \sim r_c$, $u$ and $w$ are $O(1)$, and
$$
\label{eq3083a}
\delta=(r_c-r) / r_c \approx (| \tau |-| \tau_c|) / |\tau_c|=
( | \bar{\tau} |- \bar{\tau}_c|) /
|\bar{\tau}_c | \ll 1 , \; \; \;
|\bar{\tau}| \rightarrow |\bar{\tau}_c| .
\eqno{\mbox{(3.83a)}}
$$
Far below the transition
($r \rightarrow 0$, $\delta \rightarrow 1$, $|\tau| \gg 1$,
$|\bar{\tau}| \gg |\bar{\tau}_c|$) we have
$$
\label{eq3083b}
r(\tau) \sim |\tau|^{-2} \sim
(| \bar{\tau}_c| / |\bar{\tau}|)^2 \sim
\lambda^{4/3} | \bar{\tau} |^{-2} , \; \; \;
| \bar{\tau} | \gg |\bar{\tau}_c| .
\eqno{\mbox{(3.83b)}}
$$

Let us first compare the Wulff droplet with the
various defected structures discussed in the
previous section.
According to Eqs. \req{eq3070} and \req{eq3074} the focal
conic and spherical structures are favored over the Wulff droplet
for
$\delta < q_0^{-2/3} \sim \lambda^{2/9}$, and
$\delta < q_0^{-5/6} \sim \lambda^{5/18}$, respectively.
Comparing the barriers $B_{FC} \sim B_{s ph}$ in
Eqs. \req{eq3071} and \req{eq3075} with the barrier for the defected
annulus $B_D$ in \req{eq3081}, we
see that the annulus would be favorable for
$\delta < q_0^{-1} \sim \lambda^{1/3}$, but this is precisely
the region given by Eq. \req{eq3082} where the
annulus is no longer well defined.
We conclude that it is only in a vanishingly small
region near coexistence (when the droplet size diverges and the
barrier heights are very large) that defects
come into play, and when they do it is in the form
of focal conic structures (see Fig. 10).

We are thus left with the Wulff droplet as the preferred
one over most of the metastable range of $\tau$.
In order to estimate the probability of nucleation
we need the barrier $\bar{B}_W$ in the original units of Eq. \req{eq201b},
since the lowest order estimate for this probability is the
saddle-point contribution $\exp [- \bar{B}_W]$ (the noise in
Eq. \req{eq201c} is normalized to unity).
According to Eqs. \req{eq2015h}, \req{eq2016},
\req{eq2017} and \req{eq3039}
we have
\setcounter{equation}{83}
\beq{eq3084}
\bar{B}_W = \beta B_W \sim
\lambda^{(1-d)/3} q_0^{(1-d)/2}
r^{(7-d)/4} |u|^{-1}
\tilde{B}_W (\delta) .
\eeq
Noting that $\tilde{B}_W (\delta) \sim \delta^{1-d}$ for
$\delta \ll 1$ ($r \rightarrow r_c$), and
$\tilde{B}_W (\delta) = O(1)$ for
$\delta = O(1)$, and taking into account the
asymptotic estimates in Eq. (3.83) we find
\begin{subeqnarray}
\slabel{eq3085a}
\bar{B}_W &\sim& \lambda^{(1-d)/6}
\left [ \frac{|\bar{\tau}_c|}{|\bar{\tau}| - |\bar{\tau}_c|} \right ]^{d-1}
\; \; \; |\bar{\tau}| \rightarrow |\bar{\tau}_c| , \\
\slabel{eq3085b}
\bar{B}_W &\sim& \lambda^{(1-d)/6} , \; \;\; |\bar{\tau}| /|
\bar{\tau}_c| =
O(1) , \\
\slabel{eq3085c}
\bar{B}_W &\sim& \lambda^{(5-d)/2} | \bar{\tau} |^{(d-7)/2} , \; \; \;
| \bar{\tau} | \gg | \bar{\tau}_c | \sim \lambda^{2/3} .
\end{subeqnarray}

A similar calculation for the isotropic droplet of
Eq. \req{eq3056} yields
\begin{subeqnarray}
\slabel{eq3086a}
\bar{B}_{iso} &\sim& \lambda^{(1-d)/3}
\left [ \frac{|\bar{\tau}_c|}{|\bar{\tau}| - |\bar{\tau}_c|} \right ]^{d-1}
, \; \; \;
|\bar{\tau}| \rightarrow |\bar{\tau}_c| , \\
\slabel{eq3086b}
\bar{B}_{iso} &\sim& \lambda^{(1-d)/3} , \; \;\; | \bar{\tau}| /
|\bar{\tau}_c |
= O(1), \\
\slabel{eq3086c}
\bar{B}_{iso} &\sim& \lambda^{(3-d)} |\bar{\tau}|^{d-4} , \; \; \;
|\bar{\tau}| \gg |\bar{\tau}_c| .
\end{subeqnarray}

 From the above estimates it is clear that the
anisotropic Wulff droplet has a lower barrier than
the isotropic droplet, but the barrier nevertheless is still large
[$\lambda^{(1-d)/6} \gg 1$] up to and beyond the region
$|\bar{\tau}| \approx | \bar{\tau}_c|$ when the droplet becomes
ramified according to Eq. (3.30).
Indeed, we may estimate the barrier height at
$|\bar{\tau}| \sim | \bar{\tau}_{G1} |\sim \lambda^{3/5}$,
where according to Eq. \req{eq2032} the Hartree approximation
breaks down, and we find
\beq{eq3087}
\bar{B}_W (\bar{\tau}_{G1} ) \sim \lambda^{(2-d)/5} .
\eeq
Thus for $d=3$ the Wulff barrier is still
large ($\lambda^{-1/5} \gg 1$) at this point,
which in reduced units corresponds to the
region
($|\tau| \sim |\bar{\tau}| / |\bar{\tau}_c| \sim
\lambda^{-1/15} \gg 1$).
If we continue to use the asymptotic estimate
Eq. \req{eq3085c} beyond its formal range
of validity we find that (for $d=3$) the barrier
height is of order one at
\beq{eq3088}
|\bar{\tau} | = |\bar{\tau}_{cond}| \sim
\lambda^{(5-d)/(7-d)} \sim \lambda^{1/2} ,
\eeq
which represents a crude estimate of the condensation point.
We also note from Eq. \req{eq3087} that for $d=2$ the barrier
height becomes of order unity precisely at
$\bar{\tau}= \bar{\tau}_{G1}$.

Let us compare our results with those of Fredrickson
and Binder [10].
As mentioned in the Introduction these authors confined
themselves to isotropic droplets, so we should
compare with our Eq. (3.86).
The scaled free energy \req{eq2010} of Ref [10] is the
same as our $\bar{\cal F}$, Eq. \req{eq201b}, since all quantities
are $O(1)$ except for the coupling constant
\beq{eq3088}
u \sim \bar{N}^{-1/2} \sim \lambda \ll 1
\eeq
[here and below $\lambda$ refers to our coupling constant,
not to their parameter $\lambda$, which is $O(1)$; note
also that the quantity $\delta_{FB}$ defined in their Eq. \req{eq3012}
corresponds to our
$| \bar{\tau} | - | \bar{\tau}_c |= \delta | \bar{\tau}_c|$].
The estimate of the barrier in their Eq. \req{eq3016}
is $\bar{B}_{iso} \sim \bar{N}^{-1/3}$
$[ \delta | \bar{\tau}_c| ]^{-2} \sim \lambda^{-2/3} \delta^{-2}$,
which agrees with Eq. \req{eq3086a} for
$d=3$.
However, the condition of validity of the expansion in their
Eq. \req{eq3014} is $\delta_{FB} \ll \bar{N}^{-1/3}$,
not $\delta_{FB} \ll 1$
[i.e. $\delta \ll 1$, not
$\delta \ll \lambda^{-2/3}$].
Thus the estimate of a kinetic limit of metastability
or condensation point where $\bar{B}_{iso} \sim 1$ given
by these authors,
namely $\delta_{FB} \sim \bar{N}^{-1/6} \sim \lambda^{1/3}$
(or $| \bar{\tau}_{cond}| \sim \lambda^{1/3}$) differs
significantly from our estimate
for the isotropic droplet
$| \bar{\tau}_{cond}| \sim 1$
which follows from Eq. \req{eq3086c}.
In any case, according to our calculations the critical droplet
is an anisotropic ramified structure, and it leads to the estimate
Eq. (3.88) for the kinetic limit of metastability.
\section{Conclusion}
In this paper we have presented estimates of the critical
droplet free energy of the Brazovskii model (2.1)
in the metastable phase of its fluctuation induced
first-order transition in the weak-coupling,
low-noise limit $\lambda \ll 1$.
Our work builds on that of Fredrickson and Binder [10],
but finds
a more favorable configuration
for the critical nucleus than their isotropic droplet,
by taking into account the anisotropy of the lamellar ordered
state.
Our derivation is also physically more plausible than that of
Ref. [10] in our opinion, since the free energy we use distinguishes
between the short-scale fluctuations leading to the first-order
transition and the long-wavelength fluctuations necessary to build
interfaces and droplets.
In the final results, however, this added self-consistency
changes only the numerical factors and not the basic scaling of the
barrier height in the small parameter $\lambda$.
Our main conclusion, which differs from that of
Ref. [10] is that on the scale of the first order transition
[$ |\bar{\tau}| - |\bar{\tau}_c | \sim |\bar{\tau}_c|$] the
critical barrier is large,
$\bar{B}_W \sim \lambda^{(1-d)/6} \gg 1$, for $d \geq 2$.

At this point none of our calculations can be claimed to
yield realistic estimates of the lifetimes of metastable
states in the systems described by the Brazovskii free-energy
\req{eq201b}.
This is first of all because our results are based
exclusively on dimensional analysis with no account of numerical
coefficients, and with highly idealized asymptotic conditions,
the most extreme of which is the one in Eq. \req{eq2034}.
In addition, we have only addressed the question of finding
the energy of the saddle-point configuration, which is the
first step to estimating the lifetime of a metastable state,
but by no means the whole answer.

Our work is to our knowledge the first attempt to modify classical
nucleation theory [11,12] in a controlled way to deal with
metastability in fluctuation-driven first-order transitions [13].
A straightforward extension would be to solve the self-consistent
equations \req{eq3059} \-\req{eq3061} for the Wulff
droplet numerically and thereby calculate the function
$\tilde{B}_W^{sc} ( \delta )$ explicitly.
A more difficult (and more interesting) step
would be to justify the heuristic arguments leading
to Eqs. \req{eq3042} and \req{eq3044} by a formal calculation
analogous to Langer's [12] field-theoretic derivation of the
lifetime of a metastable state in transitions where phase
coexistence already appears at the mean-field level.

The most promising application to an experimental system
is to the microphase separation transition in
symmetric diblock copolymers [6-9],
which was the main focus of the work of Fredrickson and Binder [10].
With all the caveats expressed above, we can say
that we expect the disordered phase to be metastable
against homogeneous nucleation of droplets down to a reduced
temperature of the same order as or larger than the shift
$|\bar{\tau}_c| \sim \lambda^{2/3}$ between the
mean-field and the actual (thermodynamic) transition
($|\bar{\tau}_{cond} | \geq |\bar{\tau}_c|$).
If the ordered phase is homogeneously nucleated
we expect the critical droplets to be
ramified [16], since the more regular Wulff needles shown
in Fig. 5b have large barriers.
The actual shapes that would be observed experimentally
can only be determined if one studies the subsequent
evolution of the droplets as they aggregate and coarsen [10],
a question whose elucidation requires further work.

Another possible experimental application is to
Rayleigh-B\'{e}nard convection [4,5,27,28], perhaps
near the fluid critical point where fluctuation effects
are expected to be large [29].
In that case, however, apart from non-Boussinesq effects [27] which
might mask the fluctuation contributions, it may be important
to take into account the effects of {\it multiplicative noise} [30]
on the transition, since the latter could be much larger than the additive
(thermal) noise.
It is not clear at present how much of the present theory would
be relevant in the convection system.

Acknowledgements:
One of us (PCH) wishes to acknowledge numerous
helpful discussions with David Huse.
The research of JBS was supported in part by DOE under grant
\#DE-FG03-93ER14312.
\newpage
\setcounter{section}{1}
\section*{Appendix A: Renormalization Group Recursion Relations}
\setcounter{equation}{0}
\renewcommand{\theequation}{A.\arabic{equation}}
We seek to obtain a coarse-grained free
energy by constructing a momentum-shell
renormalization group in the manner of
Wilson and Kogut [22], except that
we eliminate shells surrounding the sphere
$| {\bf q} | = q_0$, rather than shells
surrounding a single point in $q$-space,
as was done in Ref. [22].
To do this we borrow from the techniques
developed by Shankar [26] for the Fermi
liquid.
Rather than starting from the free-energy \req{eq201b},
we generalize in the usual way [22] to include
higher interactions, i.e. we start from
\beq{eqA01}
\bar{{\cal F}} [ \bar{\psi}]  =
\bar{{\cal F}}_2 +
\bar{{\cal F}}_4 +
\bar{{\cal F}}_6 + \ldots ,
\eeq
\setlength{\jot}{.2in}
\begin{subeqnarray}
\slabel{eqA02a}
\hspace*{-.4in} \bar{{\cal F}}_2 & = & \frac{1}{2}
\int_{[0, \bar{\Lambda}_0]}
d1d2 \; \bar{u}_2 (1,2)
(2 \pi )^d \delta (1+2)
\bar{\psi} (1) \bar{\psi}(2) , \\
\slabel{eqA02b}
\hspace*{-.4in} \bar{{\cal F}}_4 & = & \frac{1}{4!}
\int_{[0, \bar{\Lambda}_0 ]}
d1 \ldots d4 \;
\bar{u}_4 (1,2,3,4) \:
(2 \pi)^d \delta (1+2+3+4)
\bar{\psi}(1)
\bar{\psi}(2)
\bar{\psi}(3)
\bar{\psi}(4) ,
\end{subeqnarray}
etc, where
\beq{eqA03}
\int_{[0 ,\bar{\Lambda}_0]} \equiv
\frac{1}{(2\pi)^d}
\int_{0<| \bar{q}_1- \bar{q}_0 | < \bar{\Lambda}_0}
d \bar{q}_1 \bar{q}_1^{d-1}
\int d \Omega_1 ,
\eeq
and $d \Omega_1$ is the element of solid angle in
$d$-dimensions.
We now integrate over the modes
\beq{eqA04}
\bar{\Lambda}_0 / b < \bar{q} - \bar{q}_0 <
\bar{\Lambda}_0,
\eeq
and the corresponding inner shell with $\bar{q} < \bar{q}_0$,
and then rescale all momenta so that
$\bar{\Lambda}_0 / b \rightarrow \bar{\Lambda}_0$.
If we define
\begin{subeqnarray}
\slabel{eqA05a}
\bar{k} &=& \bar{q} - \bar{q}_0 , \\
\slabel{eqA05b}
\bar{k}^{\prime} &=& b \bar{k} , \\
\slabel{eqA05c}
\bar{\psi}^{\prime} ( \bar{k}^{\prime} ) &=&
b^{-3/2} \bar{\psi} ( \bar{k}) ,
\end{subeqnarray}
then in lowest order in $\bar{u}_4 \sim \lambda$ we may write
\beq{eqA06}
\bar{u}_2(1,2) =
[ \bar{r} + \tilde{\xi}_0^4
( \bar{q}_1^2 - \bar{q}_0^2 )^2 ] ,
\eeq
and the change in $\bar{r}$ is given by
\beq{eqA07}
\bar{r}^{\prime} =
b^2 [ \bar{r} + \Delta_2 ] ,
\eeq
where $\Delta_2$ is the contribution from the Hartree
diagram in Fig. 1a (see below).
Let us now examine the change in $\bar{u}_4$.
By an argument similar to the one given by Shankar [26] (Sec. V)
we find that at ``tree-level''
the change in $\bar{u}_4$ is given by
\beq{eqA08}
\bar{u}_4^{\prime} ( 1^{\prime} ,2^{\prime} ,
3^{\prime} , 4^{\prime} ) =
e^{-(b-1) \bar{q}_0[ | {\bf {\hat{n}}}_1 +
{\bf {\hat{n}}}_2 +
{\bf {\hat{n}}}_3 | -1 ] / \bar{\Lambda}_0}
b^3 \bar{u}_4 (1,2,3,4) ,
\eeq
where ${\bf \hat{n}}_i$ is the unit vector
${\bf \bar{q}}_i / | {\bf \bar{q}}_i |$.
Thus by iteration of the renormalization
group, only couplings with vectors satisfying
\beq{A09}
| {\bf \hat{n}}_1 +
{\bf \hat{n}}_2 +
{\bf \hat{n}}_3 | = 1
\eeq
will remain finite.
It can be shown that for both $d=2$ and $d=3$
this implies that the four wavevectors in
$\bar{u}_4(1,2,3,4)$ must be equal and opposite in
pairs when their magnitude goes to
$\bar{q}_0$.
Thus only
\beq{A010}
\bar{u}_4
( {\bf \hat{n}}_1, - {\bf \hat{n}}_1 ,
{\bf \hat{n}}_2 ,
- {\bf \hat{n}}_2 ;
\bar{q}_0 ) = \bar{u}
( {\bf \hat{n}}_1 , {\bf \hat{n}}_2)
\eeq
survives, and it satisfies
\beq{A011}
\bar{u}^{\prime}
( {\bf \hat{n}}_1 , {\bf \hat{n}}_2)
=  b^3
[ \bar{u} ( {\bf \hat{n}}_1,
{\bf \hat{n}}_2) +
\Delta_4 ( {\bf \hat{n}}_2, {\bf \hat{n}}_2 )] ,
\eeq
where $\Delta_4$ is the contribution from the diagrams in Figs. 1c and
1d.
It turns out that $\bar{u} ( {\bf \hat{n}}_1, {\bf \hat{n}}_2 )$ develops
a dependence on the angle between
${\bf \hat{n}}_1$ and ${\bf \hat{n}}_2$ under the
action of the renormalization group.
For a general nonzero angle $\bar{u}$ is practically
constant and we define
\beq{eqA012}
\bar{u} ( {\bf \hat{n}}_1, {\bf \hat{n}}_2)= \bar{u}_a .
\eeq
For the special case of parallel
${\bf \hat{n}}_1$ and ${\bf \hat{n}}_2$ (to within an
angle of $O ( \sqrt{\bar{r}}$), we define
\beq{eqA013}
\bar{u} ( {\bf \hat{n}}_1, {\bf \hat{n}}_1) = \bar{u}_b .
\eeq
Similarly, we have
\beq{A014}
\bar{u}_6 ( 1, \ldots , 6 ) \rightarrow
\bar{u}_6 ( {\bf \hat{n}}_1, - {\bf \hat{n}}_1 ,
{\bf \hat{n}}_2 , - {\bf \hat{n}}_2 ,
{\bf \hat{n}}_3 , - {\bf \hat{n}}_3 ; \bar{q}_0 )
= \bar{w} ( {\bf \hat{n}}_1 , {\bf \hat{n}}_2, {\bf \hat{n}}_3 ) ,
\eeq
\beq{A015}
\bar{w}^{\prime}
( {\bf \hat{n}}_1, {\bf \hat{n}}_2 , {\bf \hat{n}}_3 ) =
b^4 [ \bar{w}
({\bf \hat{n}}_1, {\bf \hat{n}}_2 , {\bf \hat{n}}_3) +
\Delta_6
({\bf \hat{n}}_1, {\bf \hat{n}}_2 , {\bf \hat{n}}_3) ] ,
\eeq
with $\Delta_6$ given in Figs. 1e and 1f.
Under the action of the renormalization group
$\bar{w}$ also develops an angular dependence; for
general angles between
${\bf \hat{n}}_1 , {\bf \hat{n}}_2$ and
${\bf \hat{n}}_3$
we define
\beq{eqA016}
\bar{w} = \bar{w}_a .
\eeq
When only two of the unit vectors are parallel we
define
\beq{eqA017}
\bar{w} = \bar{w}_b ,
\eeq
and for all three unit vectors parallel we define
\beq{eqA018}
\bar{w} = \bar{w}_c.
\eeq

Let us now evaluate the contributions
$\Delta_2 , \Delta_4, \Delta_6$, using the lowest
relevant order in
$\bar{u}_4 \sim \bar{u} \sim \lambda$ for each $\Delta_i$.
We find

\begin{eqnarray}
\Delta_2 & = & (2 \pi)^{-d}
\int_{\bar{q}_0+ \bar{\Lambda}_0 /b}^{\bar{q}_0+ \bar{\Lambda}_0}
d \bar{q} \; \bar{q}_0^{d-1}
\int d \Omega_1 \;
\frac{\bar{u}_4 ({\bf \hat{n}} , - {\bf \hat{n}} , {\bf \hat{n}}_1,
-{\bf \hat{n}}_1)}
{\bar{r} + \tilde{\xi}_0^4 ( \bar{q}^2 - \bar{q}_0^2 )^2} , \nonumber \\
\label{eqA019}
       & = & \bar{\alpha} \bar{\xi}_0 \bar{u}_a
\int_{\bar{\Lambda}_0 /b}^{\bar{\Lambda}_0}
d \bar{k} \;
\frac{1}{\bar{r} + \bar{\xi}_0^2 \bar{k}^2} ,
\end{eqnarray}
where $\bar{\alpha}$ is given in Eq. \req{eq2012} and $\bar{\xi}_0$ is defined
in Eq. \req{eq2015h}.
In particular there is no momentum dependence
to this order, so the second term in square brackets in
Eq. \req{eqA06} is unmodified.
In order to find differential recursion relations we set
\beq{eqA020}
b = 1 + \ell ,
\eeq
with $\ell \ll 1$, and find from Eqs. \req{eqA07} and \req{eqA019}
$$
\label{eqA021a}
\frac{d \bar{r}}{d \ell} =
2 \bar{r} +
\frac{\bar{\alpha} \bar{\xi}_0 \bar{u}_a \bar{\Lambda}_0}
{\bar{r} + \bar{\xi}_0^2 \bar{\Lambda}_0^2} .
\eqno{\mbox{(A.21a)}}
$$
The contributions to $\Delta_4$ are given
in Figs. 1c and 1d.
When we examine the contribution to $\bar{u}_a$ from
Fig. 1b we find there is one channel in which both intermediate
Green's functions have arguments in the shell being
integrated over, whereas
for $\bar{u}_b$ there are two.
This allows the evaluation of
$\Delta_4$ and leads to
$$
\label{eqA021b}
\frac{d \bar{u}_a}{d \ell} =
3 \bar{u}_a -
\frac{\bar{\alpha} \bar{\xi}_0 \bar{u}_a^2 \bar{\Lambda}_0}
{( \bar{r}+ \bar{\xi}_0^2 \bar{\Lambda}_0^2)^2} +
\frac{\bar{\alpha} \bar{\xi}_0 \bar{w}_a \bar{\Lambda}_0}
{ \bar{r} + \bar{\xi}_0^2 \bar{\Lambda}_0^2} ,
\eqno{\mbox{(A.21b)}}
$$
and
$$
\label{eqA021c}
\frac{d \bar{u}_b}{d \ell} = 3 \bar{u}_b -
\frac{2 \bar{\alpha} \bar{\xi}_0 \bar{u}_a^2 \bar{\Lambda}_0}
{ ( \bar{r} + \bar{\xi}_0^2 \bar{\Lambda}_0^2)^2} +
\frac{\bar{\alpha} \bar{\xi}_0 \bar{w}_b \bar{\Lambda}_0}
{ \bar{r} + \bar{\xi}_0^2 \bar{\Lambda}_0^2} .
\eqno{\mbox{(A.21c)}}
$$
Now consider $\Delta_6$.
There are 15 ways of distributing the 6 arguments
of $\bar{u}_6$ over the external legs in
both Figs. 1e and 1f.
Examination of the number of channels which
contribute to $\bar{w}_a , \bar{w}_b$ and $\bar{w}_c$ leads to
$$
\label{eqA021d}
\frac{d \bar{w}_a}{d \ell} =
4 \bar{w}_a +
\frac{2 \bar{\alpha} \bar{\xi}_0 \bar{u}_a^3 \bar{\Lambda}_0}
{ ( \bar{r} + \bar{\xi}_0^2 \bar{\Lambda}_0^2 )^3} -
\frac{3 \bar{\alpha} \bar{\xi}_0 \bar{u}_a \bar{w}_a \bar{\Lambda}_0}
{( \bar{r} + \bar{\xi}_0^2 \bar{\Lambda}_0^2)^2} ,
\eqno{\mbox{(A.21d)}}
$$
and two similar equations for $\bar{w}_b$ and $\bar{w}_c$.

These equations have initial values corresponding to Eq. \req{eq201b},
\setcounter{equation}{21}
\begin{subeqnarray}
\slabel{eqA022a}
\bar{r} ( \ell = 0) & = & \bar{\tau} , \\
\slabel{eqA022b}
\bar{u}_a ( \ell =0) =
\bar{u}_b ( \ell =0) & = &\lambda , \\
\slabel{eqA022c}
\bar{w}_a ( \ell =0) =
\bar{w}_b ( \ell =0) =
\bar{w}_c ( \ell = 0 ) & = & 0 .
\end{subeqnarray}
Because of the scalings we have chosen in Eq. (A.5) the
quantities $\bar{r} ( \ell)$, $\bar{u}_{a,b} ( \ell)$,
$\bar{w}_{a,b,c} ( \ell)$ all grow with $\ell$.
To find finite quantities we make a change of variables
\begin{subeqnarray}
\slabel{eqA023a}
\Lambda &=& \bar{\Lambda}_0 \bar{\xi}_0 ( \bar{\alpha} \lambda)^{-1/3}
e^{-\ell} , \\
\slabel{eqA023b}
r ( \Lambda) &=& \bar{r} (\ell)
(\bar{\alpha} \lambda)^{-2/3} e^{-2\ell} , \\
\slabel{eqA023c}
u_{a,b} ( \Lambda) &=& \bar{u}_{a,b} ( \ell)
\lambda e^{-3\ell} , \\
\slabel{eqA023d}
w_{a,b,c} ( \Lambda ) &=& \bar{w}_{a,b,c} ( \ell )
\bar{\alpha}^2 ( \bar{\alpha} \lambda)^{-4/3}
e^{-4 \ell} ,
\end{subeqnarray}
leading to
\begin{subeqnarray}
\slabel{eqA024a}
\frac{dr}{d \Lambda} &=& -
\frac{u_a}{r+\Lambda^2} , \\
\slabel{eqA024b}
\frac{du_a}{d\Lambda} &=&
\frac{u_a^2}{(r+\Lambda^2)^2} -
\frac{w_a}{r+\Lambda} , \\
\slabel{eqA024c}
\frac{du_b}{d \Lambda} &=& \frac{2u_a^2}{(r+\Lambda^2)^2} -
\frac{w_b}{r+\Lambda^2} , \\
\slabel{eqA024d}
\frac{dw_a}{d\Lambda} &=&
\frac{-2u_a^3}{(r+ \Lambda^2)^3} +
\frac{3u_aw_a}{(r+\Lambda^2)^2} ,\\
\slabel{eqA024e}
\frac{dw_b}{d \Lambda} &=&
\frac{-4u_a^3}{(r+\Lambda^2)^3} +
\frac{u_aw_b}{(r+\Lambda^2)^2} +
\frac{4u_aw_a}{(r+\Lambda^2)^2} , \\
\slabel{eqA024f}
\frac{dw_c}{d \Lambda} &=& \frac{-12u_a^3}{(r+ \Lambda^2)^3} +
\frac{8u_aw_b}{(r+\Lambda^2)^2},
\end{subeqnarray}
with initial values at
$\Lambda = \bar{\Lambda}_0 \bar{\xi}_0 ( \bar{\alpha} \lambda)^{-1/3} =
\infty$,
$$
\label{eqA025a,b,c}
r( \infty ) = \tau, \; \; \; \;
u_{a,b} (\infty) = 1, \; \; \; \;
w_{a,b,c}  ( \infty) = 0 .
\eqno{\mbox{(A.25a,b,c)}}
$$

Let us reexpress the free energy in terms of the running couplings
$\bar{u}_n ( \ell)$.
Since the upper cutoff $\bar{\Lambda}_0$ is restored to its
original value at each stage of the recursion \req{eqA07},
the effective free energy still has the form (A.1-2) when
expressed in terms of the $\bar{u}_n( \ell)$ and the corresponding
$\bar{\psi} ( \ell)$.
Let us now change to the scaling (A.23), supplemented
with the relation
\setcounter{equation}{25}
\beq{eqA026}
\psi ( \ell) =
\bar{\psi} ( \ell)
\lambda^{1/2} ( \bar{\alpha} \lambda)^{-1/3}
e^{3\ell /2} .
\eeq
Then it is straightforward to show that the free energy
takes precisely the form (2.44), with a variable upper cutoff $\Lambda$,
and coefficients satisfying Eqs. (A.24).
The resulting functions
$r( \Lambda), u ( \Lambda ) = u_b (\Lambda), w ( \Lambda )$
$= w_c(\Lambda)$ are similar, but not identical
to the ones defined in Eqs. (2.45-48) and obtained phenomenologically
in Appendix B.
In particular, contrary to the latter functions, the
solutions of (A.24) do not exist for all $\tau$ and
$\Lambda$, due to singular denominators for $\tau < \tau_1 = -2.66$.
This is to be contrasted with the bulk transition
predicted by (A.24) at
$\tau_c^{rg} = - 2.56$.
An analysis of Eqs. (A.24) shows that the
coefficients are undefined for
\beq{eqA027}
\Lambda < \Lambda_1 ( \tau ) ,
\eeq
where $\Lambda_1 ( \tau)$ is plotted in Fig. 11.
In the domain
\req{eqA027} represented by the shaded region of Fig. 11,
the quantity $r (\Lambda, \tau ) + \Lambda^2$ appearing in the
denominators in (A.24) has zeroes, so the recursion relations cannot be
solved with initial conditions at $\Lambda = \infty$.
Of course the singularities in
(A.24) occur for $r(\Lambda,\tau) < 0$, i.e. in
a region where the disordered state is not even metastable,
so they are not physically significant.
However, in the renormalization group
formulation the coarse-grained free energy at small $\Lambda$
is found by integration starting from a bare free energy
with large $\Lambda$, so the recursion relations are needed even
in the unphysical parameter range with
$\tau < \tau_c$, $\Lambda$ large, and
$r < 0$.
The phenomenological coarse-graining procedure of
Eqs. (2.45-48) avoids the singularity,
but even if it did not, this would not
affect the utility of the method.
This is because the phenomenological equations
are given in integrated form,
so the coefficients $r ( \Lambda)$, $u(\Lambda)$, $w ( \Lambda)$
at a particular value of $\Lambda$ are obtained by solving equations
such as (2.45-48) involving only the same value of $\Lambda$,
rather than by integrating down from large $\Lambda$.
This means that the physically relevant small-$\Lambda$
behavior can be obtained independently of any large-$\Lambda$
singularities.

In Fig. 12 we show the numerical solution of the recursion
relations (A.24) in a form similar to that in Fig. 2.
The results only exist to the right of the shaded region
in Fig. 11, i.e. for $\tau > \tau_1$
when $\Lambda =0$.
The coefficients agree with the phenomenological
ones for $\tau \gtrsim 0$, but they differ
quantitatively for $\tau < 0$ due to the
vanishing of $r$ at $\tau_1$.
Nevertheless, for $\tau > \tau_1$ the recursion relations provide
a justification for the
phenomenological theory,
since they have been derived by more or less
standard diagrammatic methods.
\newpage
\section*{Appendix B: Phenomenological Coarse-graining Procedure}
\setcounter{equation}{0}
\renewcommand{\theequation}{B.\arabic{equation}}
In this Appendix we reformulate the
derivation of Fredrickson and Binder [10] for
the effective free energy, replacing the averages in their
Eq. \req{eqA03}, which are over the whole range
$0 < | q - q_0 | < \Lambda_0$, by
averages over the restricted range
$0 < | q-q_0 | < \Lambda$.
The coefficients $\bar{r}(\Lambda)$,
$\bar{u}(\Lambda)$, $\bar{w} ( \Lambda)$ are
just the derivatives
\beq{eqB01}
\Gamma_{\Lambda n} =
\frac{\delta^n \bar{{\cal F}}_{\Lambda}}{\delta \bar{\psi}^n} ,
\eeq
for $n= 2,4,6$, respectively.
They are obtained from the diagrams in Fig. 1
except that the integrals are confined to
the range
$0 < | q-q_0| < \Lambda$,
and no rescaling of the momentum is
performed.
As we noted in our discussion of the renormalization group in
Appendix A, the number of channels which contribute to
$\Gamma_4$ and $\Gamma_6$ depends on the angles between
the wavevectors which are their arguments.
We will eventually want the free energy to be a functional of
a $\psi$ of the form \req{eq303}.
Thus we want the wavevectors occurring in the arguments
to be either parallel or antiparallel.
In the notation of Appendix A we then want to identify
$u_b$ with $u$ and $w_c$ with $w$.

For $r$ the Hartree graph in Fig. 1a gives
\beq{eqB02}
\bar{r} ( \Lambda ) =
\bar{\tau} + \bar{\alpha} \bar{\xi}_0 \lambda
\int_{\bar{\Lambda}}^{\bar{\Lambda}_0}
\frac{d \bar{k}}{\bar{r}+ \bar{\xi}_0^2 \bar{k}^2}.
\eeq
For $\bar{u}_a$ the graph in Fig. 1c yields
\beq{eqB03}
\bar{u}_a = \lambda -
\bar{u}_a \lambda
\int_{\bar{\Lambda}}^{\bar{\Lambda}_0} \bar{\alpha}
\bar{\xi}_0
\frac{d \bar{k}}{( \bar{r}+ \bar{\xi}_0^2 \bar{k}^2)^2} .
\eeq
For $\bar{u}_b$ this graph yields
\beq{eqB04}
\bar{u}_b = \lambda
-2 \bar{u}_a \lambda
\int_{\bar{\Lambda}}^{\bar{\Lambda}_0} \bar{\alpha}
\bar{\xi}_0
\frac{d \bar{k}}{(\bar{r}+ \bar{\xi}_0^2 \bar{k}^2)^2} .
\eeq
The coefficient of the sixth order term, $\bar{w}_c$, is
given by the graph in Fig. 1e which yields
\beq{eqB05}
\bar{w}_c = 12
\int_{\bar{\Lambda}}^{\bar{\Lambda}_0} \bar{u}_a^3 \bar{\alpha}
\bar{\xi}_0
\frac{dk}{(\bar{r}+ \bar{\xi}_0^2 \bar{k}^2)^3} .
\eeq
When Eqs. (B.2-B.5) are reexpressed in terms of the
scaled units of Eq. (2.15) they then yield precisely
Eqs. (2.45-48).

Let us verify that the denominator in the
$\phi_n$, Eq \req{eq2048} remains positive
\beq{eqB06}
r(\Lambda, \tau )+ \Lambda^2 > 0 ,
\eeq
from which it follows that the coefficients
$r(\Lambda, \tau )$,
$u (\Lambda , \tau)$ and
$w ( \Lambda, \tau)$, Eqs. (2.45-48) are
well-defined for all $\Lambda$ and $\tau$.
The question only arises for
$r <0$ and $\tau < 0$, so we set
\beq{eqB07}
r=- \Lambda^2 (1- \eta )^2 ,
\eeq
and ask whether $\eta$ can vanish.
If we carry out the integral $\phi_1$ in Eq. \req{eq2048} we
find (assuming $\eta > 0$)
\beq{eqB08}
r=
- \Lambda^2 (1-\eta)^2 =
\tau + \frac{1}{2\Lambda(1-\eta)}
\ln \left ( \frac{2-\eta}{\eta} \right ) ,
\eeq
which for $\eta \rightarrow 0$ becomes
\beq{eqB09}
\ln \frac{2}{\eta} =
2 \Lambda ( | \tau | - \Lambda^2) .
\eeq
Since for fixed $\tau$ the rhs of \req{eqB09} is
bounded
by $4 ( | \tau | /3)^{3/2}$
as a function of $\Lambda$, we conclude that \req{eqB09}
has no solution for $\eta \rightarrow 0$, and that the
inequality \req{eqB06} is always satisfied.
\clearpage
\section*{References}
\bigskip
\begin{enumerate}
\item
S. A. Brazovskii, Sov. Phys. JETP {\bf 41}, 85 (1975).
\item
D. Briskin, D. L. Johnson, H. Fellner, and
M. E. Neubert,
Phys. Rev. Let. {\bf 50}, 178 (1983);
L. J. Martinez-Miranda, A. R. Kortan,
and R. J. Birgeneau, {\it ibid} {\bf 56},
2264 (1986);
C. W. Garland and M. E. Huster,
Phys. Rev. A{\bf 35}, 2365 (1987).
\item
R. F. Sawyer, Phys. Rev. Lett. {\bf 29}, 382 (1972);
A. N. Dyugaev, JETP Lett. {\bf 22}, 83 (1975).
\item
J. B. Swift and P. C. Hohenberg,
Phys. Rev. A{\bf 15}, 319 (1977).
\item
P. C. Hohenberg and J. B. Swift,
Phys. Rev. A{\bf 46}, 4773 (1992).
\item
L. Leibler, Macromolecules {\bf 13}, 1602 (1980).
\item
G. H. Fredrickson and E. Helfand, J. Chem. Phys. {\bf 87}, 697 (1987).
\item
F. S. Bates, J. H. Rosedale, G. H. Fredrickson and C. J. Glinka,
Phys. Rev. Lett. {\bf 61}, 2229 (1988).
\item
F. S. Bates, J. H. Rosedale, and G. H. Fredrickson,
J. Chem. Phys. {\bf 92}, 6225 (1990).
\item
G. H. Fredrickson and K. Binder,
J. Chem. Phys. {\bf 91}, 7265 (1989).
\item
F. F. Abraham, {\it Homogeneous Nucleation Theory}
(Academic, NY, 1974); J. D. Gunton, M. San Miguel, and P. S. Sahni,
in {\it Phase Transitions and Critical Phenomena},
edited by C. Domb and J. L. Lebowitz (Academic, NY, 1983)
Vol. 8; J. D. Gunton and M. Droz,
{\it Introduction to the Theory of Metastable and Unstable States},
Lecture Notes in Physics 183 (Springer Verlag, Berlin, 1983).
\item
J. S. Langer, Ann. Phys. (NY) {\bf 41},
108 (1967); {\bf 54}, 258 (1969);
in {\it Systems Far from Equilibrium}, Lecture Notes in Physics
132 (Springer Verlag, Berlin, 1980).
\item
J. Rudnick, Phys. Rev. B{\bf 18}, 1406 (1978)
and references therein.
\item
See L. D. Landau and E. M. Lifshitz,
{\it Statistical Physics} (Addison-Wesley, London, 1958) Ch. XV.
\item
J. B. Fournier and G. Durand, J. Phys. II France {\bf 1},
845 (1991).
\item
C. Unger and W. Klein, Phys. Rev. B{\bf 29}, 2698 (1984).
\item
P. C. Hohenberg and B. I. Halperin, Rev. Mod. Phys. {\bf 49}, 435
(1977).
\item
V. L. Ginzburg, Sov. Phys. Sol. State {\bf 2}, 1824 (1960).
\item
For greater clarity of presentation we have given all our
definitions in this subsection in terms of the unbarred (scaled)
quantities $q , \Lambda , \psi , {\cal F}$, etc of
Eqs. (2.15), but analogous definitions would
hold for the original barred quantities.
In Appendix A we define a different coarse graining
procedure in terms of quantities $\bar{r} ( \bar{\Lambda})$,
$\bar{u} ( \bar{\Lambda})$,
$\bar{w} ( \bar{\Lambda})$ which are {\it not}
obtained from $r(\Lambda)$, $u( \Lambda)$ $w ( \Lambda)$ via
Eq. (2.15).
\item
Since $\Lambda_0=q_0$, and
according to Eq. (2.15)
$q_0=(2 \tilde{\xi}_0^2)^{-1}$
$( \alpha \lambda )^{-1/3} \gg 1$, for
$\lambda \ll 1$, we will henceforth set
$\Lambda_0= \infty$, as in the upper limit of the integral of
Eq. \req{eq2048}.
\item
In this paper we refer to the function $\bar{{\cal F}}$ in
the starting Brazovskii model \req{eq201b} as a ``free energy''
since it is obtained by coarse-graining a more microscopic
Hamiltonian.
Subsequent coarse-grained versions, denoted ${\cal F}_{\Lambda}$
are also referred to as ``free energies''.
 From the point of view of nucleation theory, on the other hand,
since we are only discussing saddle-point configurations, the
quantities we calculate are ``energy barriers''.
In order to estimate ``free-energy barriers'', we would
have to take into account the ``entropic'' contributions to
the lifetime of the metastable state coming from
other configurations, for example the small oscillations
around the saddle.
Since we neglect these effects in the present work we use
the terms energy and free energy somewhat loosely.
\item
K. G. Wilson and J. B. Kogut, Phys. Rept. {\bf 12}, 7 (1974).
\item
S. K. Ma, {\it Modern Theory of Critical Phenomena}
(Benjamin-Cummings, Reading, MA, 1976).
\item
D. Mukamel and R. M. Hornreich, J. Phys. C{\bf 13}, 161 (1980).
\item
D. D. Ling, B. Friman, and G. Grinstein,
Phys. Rev. B{\bf 24}, 2718 (1981).
\item
R. Shankar, Rev. Mod. Phys. {\bf 66}, 129 (1994).
\item
See M. C. Cross and P. C. Hohenberg, Rev. Mod. Phys. {\bf 65},
851 (1993), Sec. IV.A.1.
\item
M. C. Cross, Phys. Rev. A{\bf 25}, 1065 (1982).
\item
V. Steinberg and S. A. Brazovskii, (unpublished).
\item
M. V. Feigelman and I. E. Staroselsky,
Z. Phys. B. Cond. Mat. {\bf 62}, 261 (1986).
\end{enumerate}
\newpage
\section*{Figure Captions}
\bigskip
\begin{enumerate}
\item
Low-order diagrams in the weak-coupling
expansion of the Brazovskii model \req{eq201b}.
\begin{itemize}
\item[(a)]
Hartree diagram contributing to $r$, or $\Delta_2$.
\item[(b)]
The second-order self-energy diagram entering
$\Sigma_2$ in Eq. \req{eq2029}.
\item[(c)]
A diagram contributing to $\Delta_4$.
The three ways of distributing the arguments
of $u_4(1,2,3,4)$ are shown.
\item[(d)]
Another diagram contributing to $\Delta_4$.
\item[(e)]
A diagram contributing to $\Delta_6$.
The fifteen ways of distributing the arguments of
$u_6(1,2,3,4,5,6)$ are not shown.
\item[(f)]
Another diagram contributing to $\Delta_6$.
\end{itemize}
\item
The coefficients of the bulk free energy \req{eq2023}, plotted
as a function of $\tau$.
(a) The coefficient $r$, obtained by solving Eq. \req{eq2019}.
(b) The coefficient $u$, from Eq. \req{eq2024}.
(c) The coefficient $w$, from Eq. \req{eq2025}.
The bulk coefficients correspond to the limit
$\Lambda =0$ of the coefficients of Eqs. (2.45-48).
\item
Schematic diagram of a longitudinal interface
between the lamellar ($x > 0$) and disordered
state ($x < 0$).
(a) Sketch of the lamellae; (b) Order parameter
$|A|$ vs. $x$.
The width of the interface is $\xi_{\parallel}$.
\item
As in the preceding figure, but for a transverse interface,
with lamellae for $y > 0$ and a disordered state for
$y<0$.
The transverse interface is thinner than the longitudinal one
($\xi_{\perp} \ll \xi_{\parallel}$).
\item
Schematic diagram of critical droplets.
(a) Isotropic droplet with concentric lamellae.
(b) The anisotropic Wulff droplet, consisting of lamellae
perpendicular to $x$.
\item
The parallelipiped entering the Wulff construction.
\item
A portion of the Brazovskii sphere in reciprocal space, showing the
relationship between the longitudinal cutoff $\Lambda$ and
the transverse cutoff $( 2q_0 \Lambda)^{1/2} \gg \Lambda$,
for $\Lambda \ll q_0$.
\item
Schematic representation of focal conic defects.
(a) and (b) represent a single focal conic
introduced into the Wulff droplet; (c) is a
biconical domain in which focal conics are juxtaposed;
(d) is a focal conic spherical network of cones representing the
equilibrium shape for arbitrarily large droplets.
[From Ref. 15.]
\item
Possible distortions of the Wulff droplet to
eliminate longitudinal interfaces.
(a) An annular shape with no defects;
(b) the same, but with the bend relieved by
the introduction of dislocations.
\item
A summary of the metastability properties of the
Brazovskii model in three dimensions, for $\lambda \ll 1$.
Shown are both the scale of the control parameter $\bar{\tau}$
of Eq. \req{eq201b}, and of the reduced quantity
$\tau \sim \bar{\tau} \lambda^{-2/3}$.
The fluctuation induced bulk
transition occurs for
$\tau < 0$, at
$\tau = \tau_c = O(1)$
$[ \bar{\tau} = \bar{\tau}_c =O( \lambda^{2/3})]$.
The defected critical droplets are only favorable
in an infinitesimal region near $\tau_c$,
$\delta = [ | \tau |- |\tau_c|] / |\tau_c| \sim \lambda^{2/9} \ll 1$.
The anisotropic Wulff droplet is favorable for
$1 > \delta > \lambda^{2/9}$.
For $|\tau| \gg |\tau_c|$,
$[|\bar{\tau}| \gg |\bar{\tau}_c|]$ the Wulff
droplet is ramified since the interface width is of the same order
as the droplet radius.
The Ginzburg criterion beyond which the perturbation
theory no longer holds occurs for
$|\tau|= |\tau_{G1}| \sim \lambda^{-1/15} \gg |\tau_c|$,
[$|\bar{\tau}|= |\bar{\tau}_{G1}| = \lambda^{3/5} \gg |\bar{\tau}_c|$], at
which point the critical barrier height is
$\bar{B}_{G1} \sim \lambda^{-1/5} \gg 1$.
The barrier becomes of order unity at
$|\tau| = |\tau_{cond}| \sim \lambda^{-1/6}$,
$[|\bar{\tau}|= |\bar{\tau}_{cond}| \sim \lambda^{1/2}]$.
\item
The function $\Lambda_1(\tau)$ at which the recursion
relations (A.24) have a singularity.
The coefficients are well-defined for all $\Lambda$
if $\tau > \tau_1$ and for $\Lambda > \Lambda_1(\tau)$ if
$\tau < \tau_1$.
\item
The bulk coefficients obtained from the recursion
relations (A.24). (a) The coefficient
$r( \Lambda =0 , \tau )$ vs. $\tau$;
(b) $u(\tau) = u_b ( \Lambda =0 , \tau )$
vs. $\tau$; (c) $w (\tau) = w_c( \Lambda =0, \tau)$ vs. $\tau$.
The coefficients are only defined for $\tau > \tau_1 =-2.65$.
These results are to be compared with those in Fig. 2.
\end{enumerate}
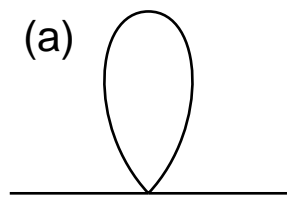
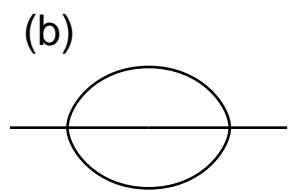
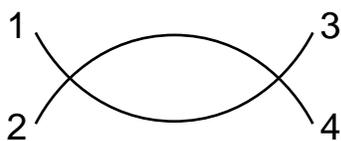
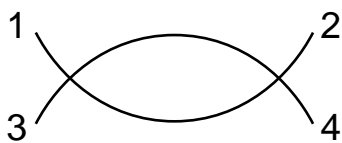
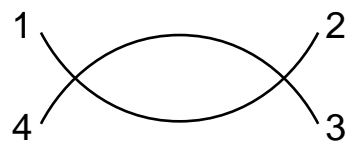
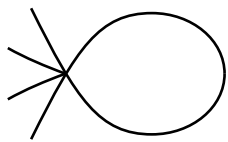
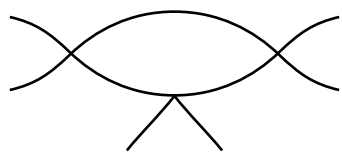
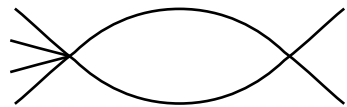
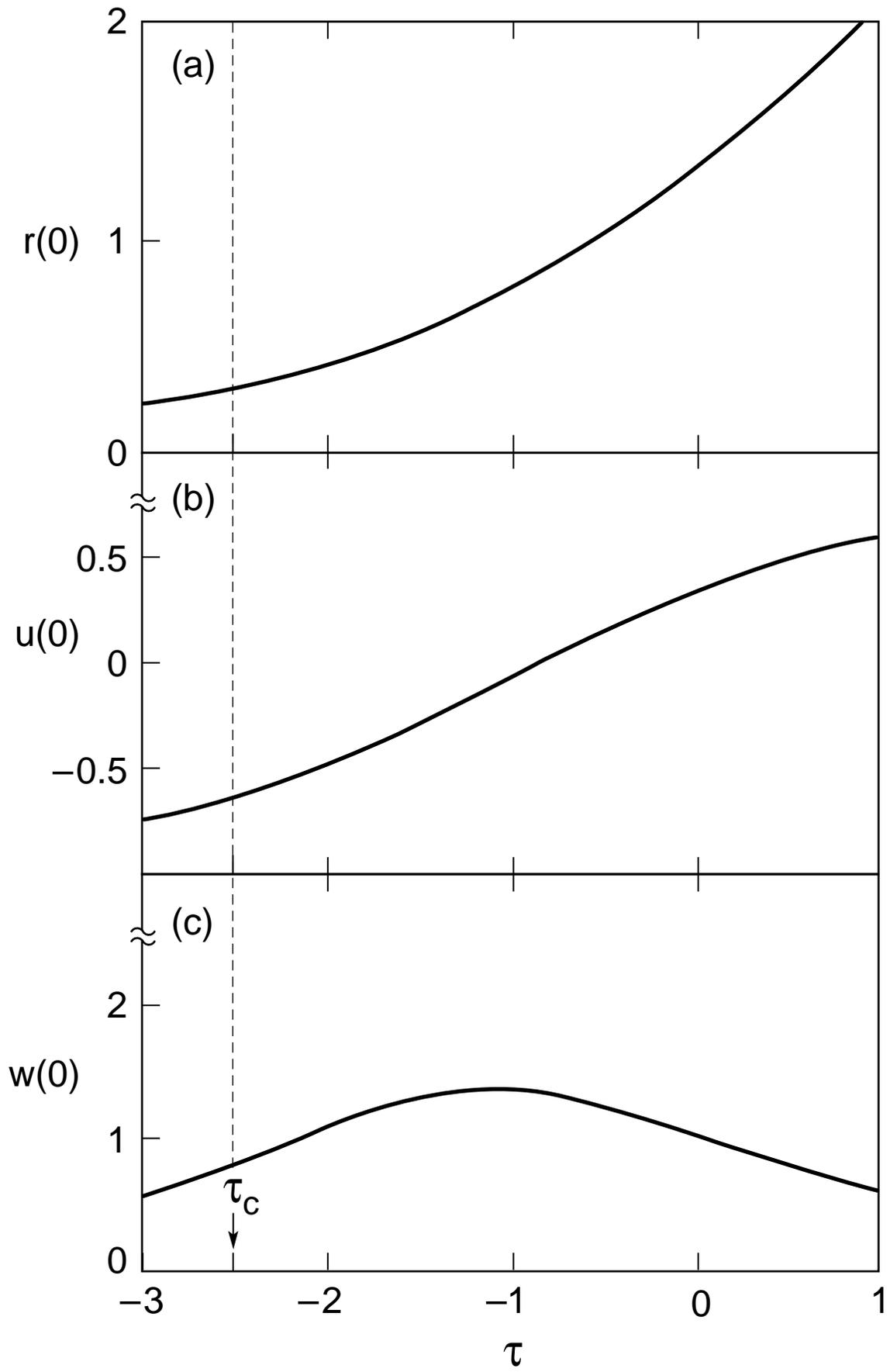
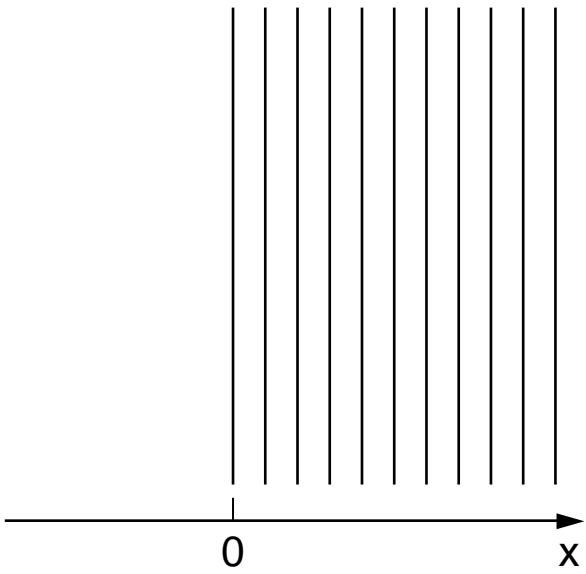
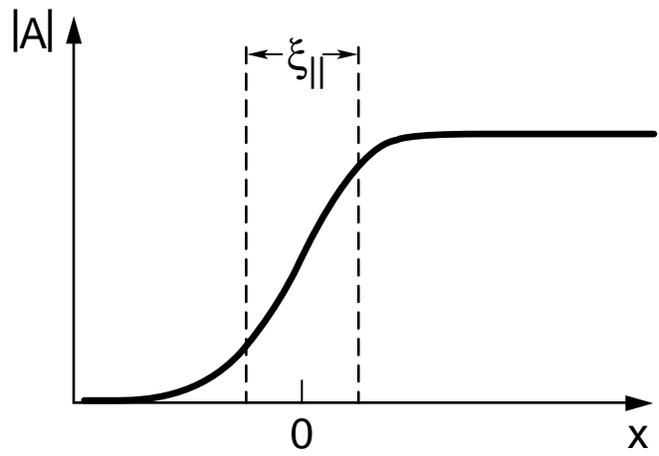
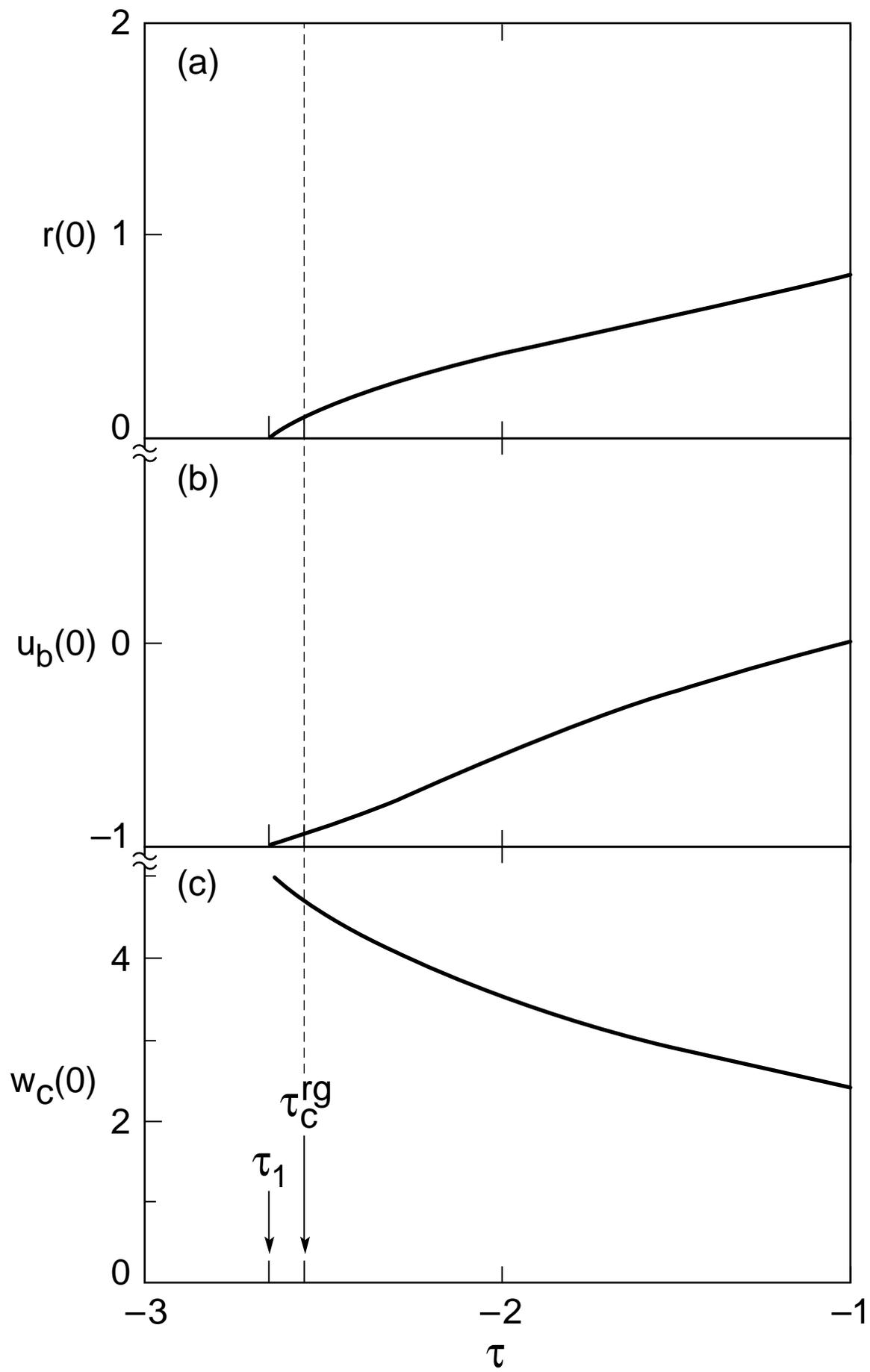
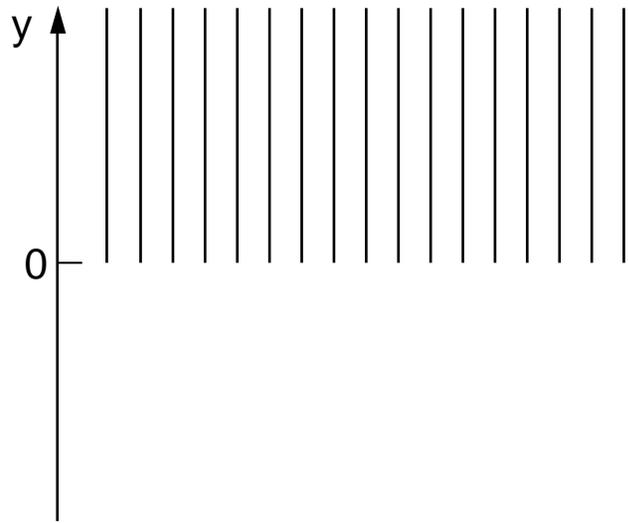
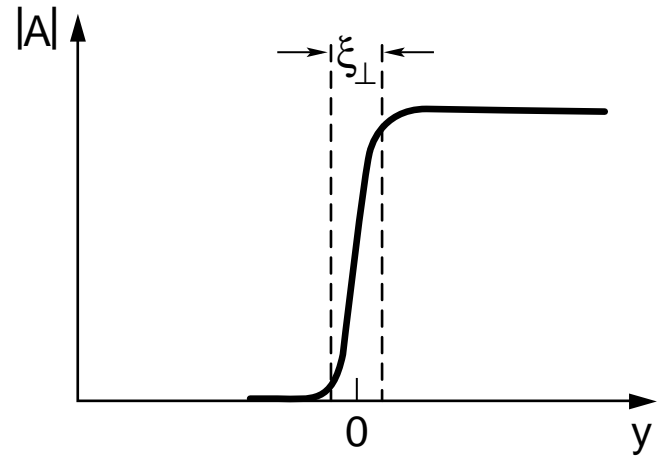
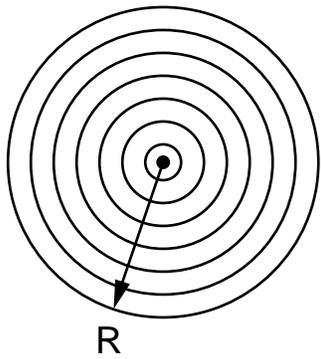
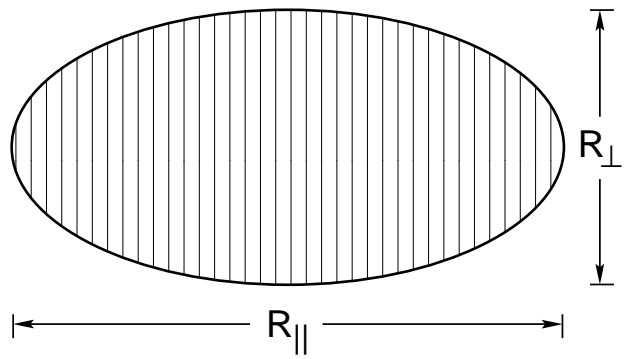
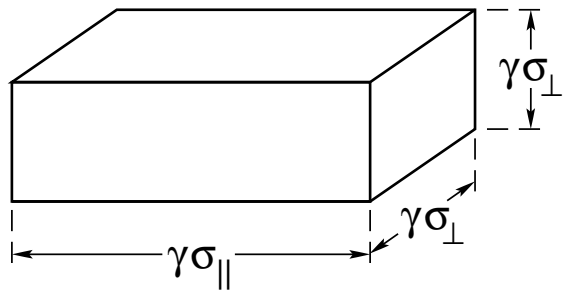
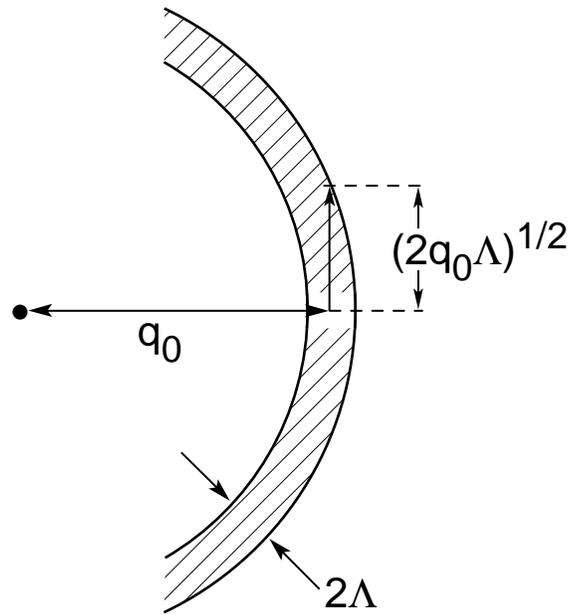
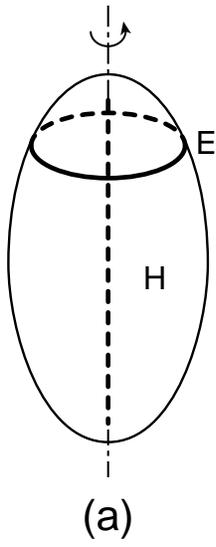
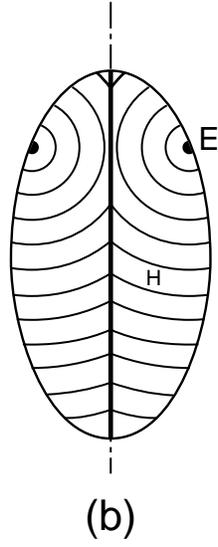
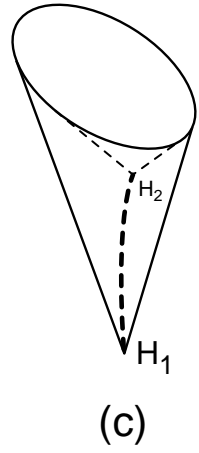
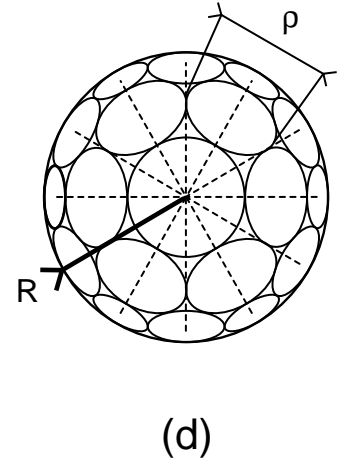
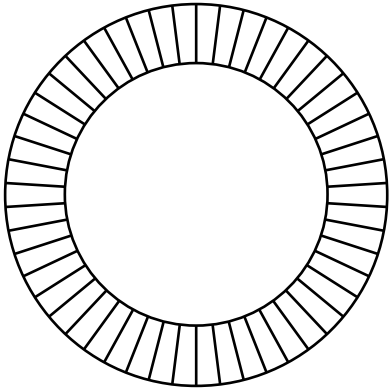
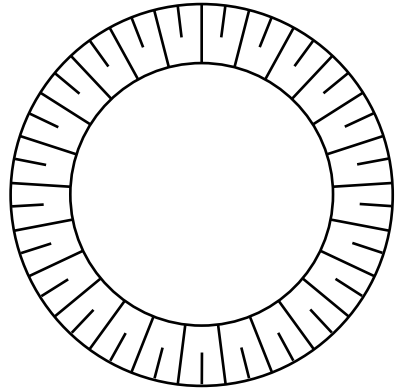
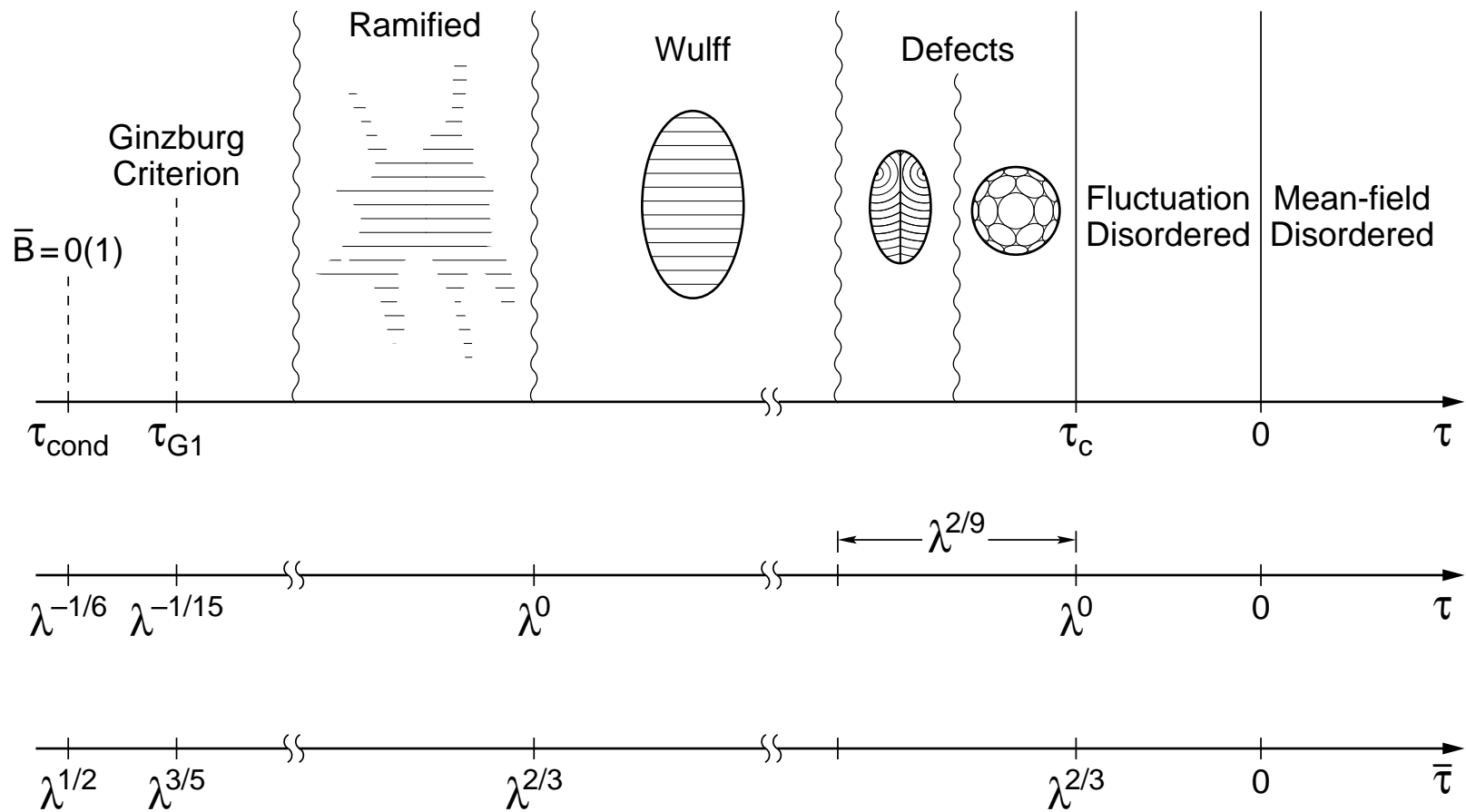
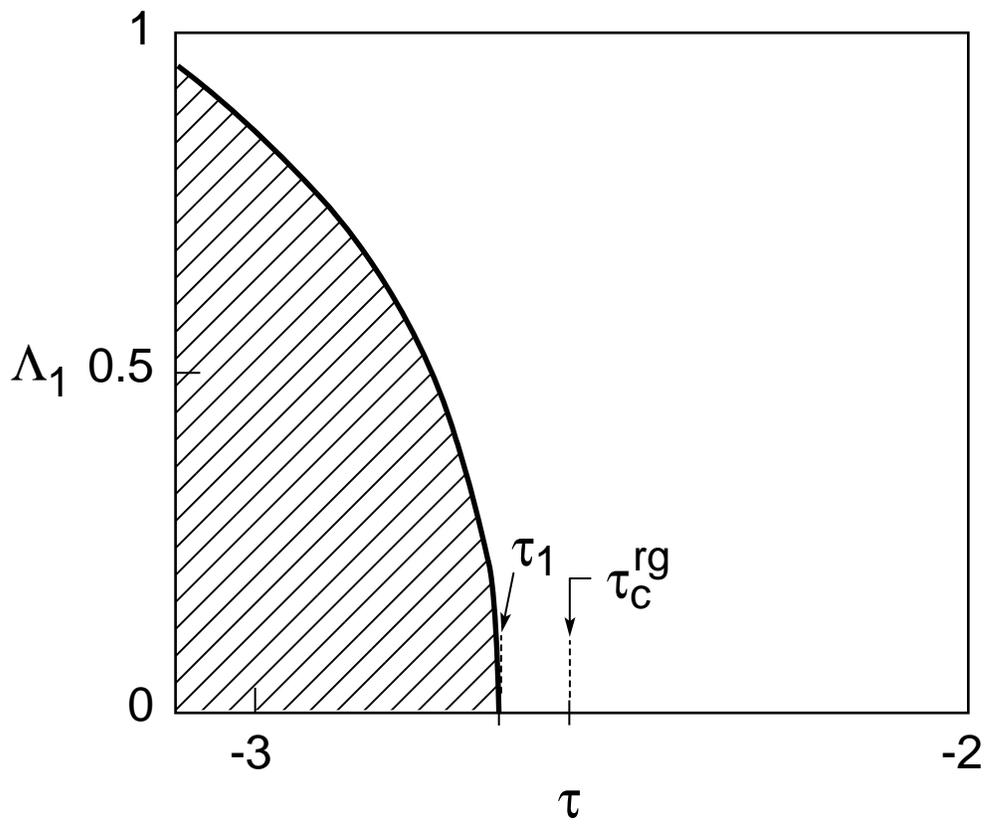
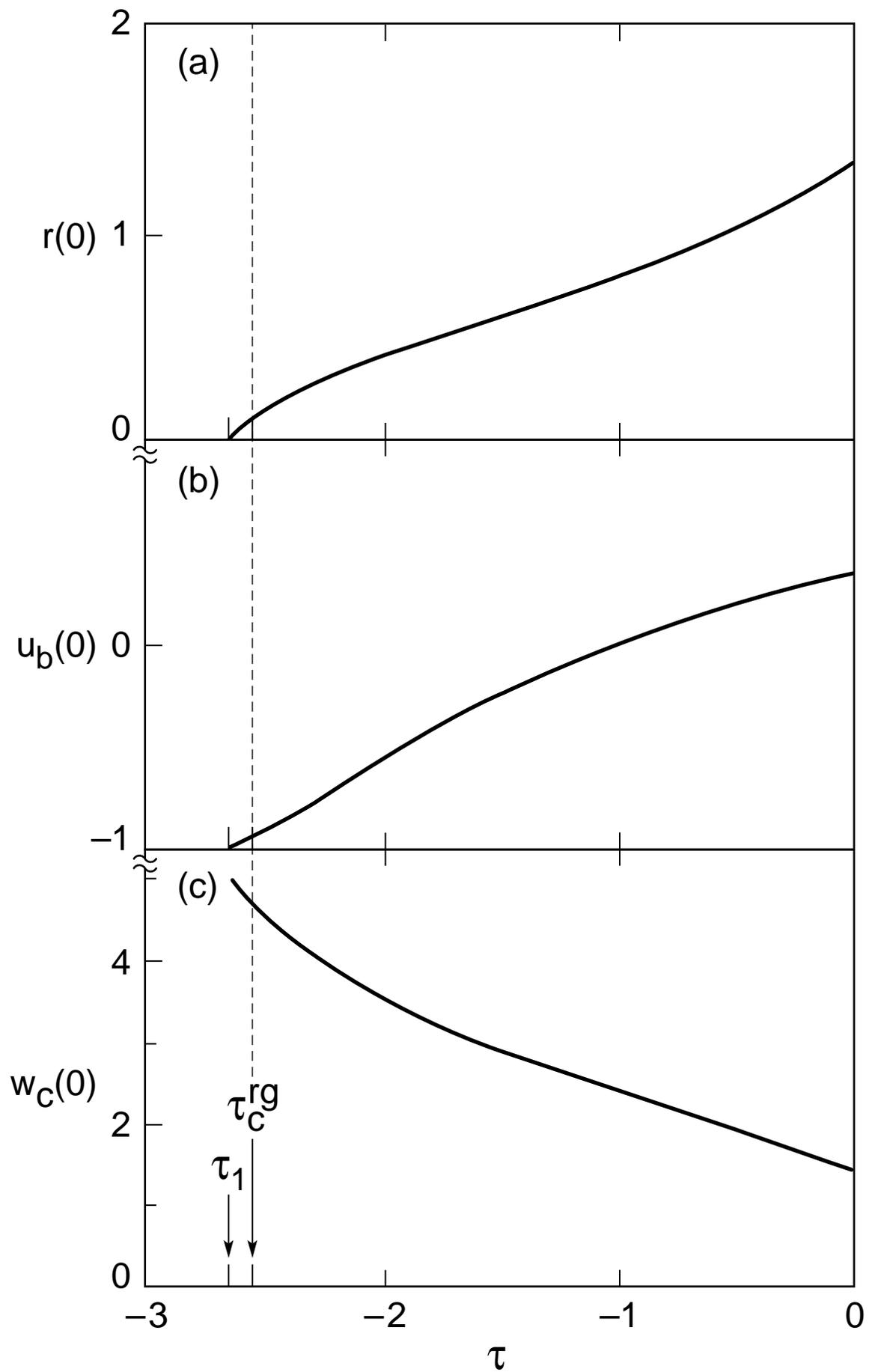
\end{document}